\documentclass[twocolumn]{revtex4}
\usepackage[dvips]{graphicx}
\usepackage{float, color,array, multirow}

\begin{document}

\title{Hedgehog lattice and field-induced chirality in breathing-pyrochlore Heisenberg antiferromagnets}

\author{Kazushi Aoyama$^1$ and Hikaru Kawamura$^2$}

\date{\today}

\affiliation{${}^{1}$Department of Earth and Space Science, Graduate School of Science, Osaka University, Osaka 560-0043, Japan \\
${}^{2}$Molecular Photoscience Research Center, Kobe University, Kobe 657-8501, Japan
}

\begin{abstract}
We theoretically investigate a $J_1$-$J_3$ classical Heisenberg model on the breathing pyrochlore lattice, where the nearest-neighbor (NN) exchange interactions for small and large tetrahedra, $J_1$ and $J_1'$, take different values due to the breathing bond-alternation and $J_3$ is the third NN antiferromagnetic interaction along the bond direction. It is found by means of Monte Carlo simulations that for large $J_3$, a hedgehog lattice, a three-dimensional periodic array of magnetic monopoles and antimonopoles, emerges in the form of a quadruple-${\bf Q}$ state characterized by the ordering vector of ${\bf Q}=(\pm\frac{1}{2},\pm\frac{1}{2},\pm\frac{1}{2})$, being irrespective of the signs of $J_1$ and/or $J_1'$ as long as $J_1\neq J_1'$. It is also found that in an applied magnetic field, there appear six quadruple-${\bf Q}$ states depending on the values of $J_1$ and $J_1'$, among which three phases including the in-field hedgehog-lattice state exhibit nonzero total chirality $\mbox{\boldmath $\chi$}^{\rm T}$ associated with the anomalous Hall effect of chirality origin. In the remaining two chiral phases, which are realized in the presence of ferromagnetic $J_1$ and/or $J_1'$, the spin structure is not topologically nontrivial, in spite of the fact that $\mbox{\boldmath $\chi$}^{\rm T} \neq 0$. The role of the topological objects of the monopoles in $\mbox{\boldmath $\chi$}^{\rm T}$ is also discussed.
\end{abstract}

\maketitle
\section{introduction}
During the last decades, a magnetic skyrmion and its two-dimensional periodic array, a skyrmion lattice, have extensively been studied in noncentrosymmetric magnets with the Dzaloshinskii-Moriya (DM) interaction \cite{SkX_review_Nagaosa-Tokura_13, MnSi_Kadowaki_82, SkX_Yi_09, SkX_Buhrandt_13, MnSi_Muhlbauer_09, MnSi_Neubauer_09, FeCoSi_Yu_10, FeGe_Yu_11, Fefilm_Heinze_11, Cu2OSeO3_Seki_12, CoZnMn_Tokunaga_15, GaV4S8_Kezsmarki_15, GaV4Se8_Fujima_17, GaV4Se8_Bordacs_17, VOSe2O5_Kurumaji_17, AntiSkX_Nayak_17, EuPtSi_Kakihana_18, EuPtSi_Kaneko_19} and recently, DM-free centrosymmetric magnets as well \cite{SkX_Okubo_12, SkX_Leonov_15, SkX_Lin_16, SkX_Hayami_16, SkXimp_Hayami_16, SkX_top2_Ozawa_prl_17, SkX-RKKY_Hayami_17, SkX_Lin_18, SkX-RKKY_Hayami_19, SkX-RKKY_Wang_20, SkX-bondaniso_Hayami_21, SkX-bondaniso_Batista_21, SkX-RKKY_Mitsumoto_21, SkX-RKKY_Mitsumoto_22, Gd2PdSi3_Kurumaji_19, GdRuAl_Hirschberger_natcom_19, GdRu2Si2_Khanh_20}. Compared to the two-dimensional skyrmion, less is known about an associated three-dimensional topological spin texture, a magnetic hedgehog, and its periodic array, a hedgehog lattice \cite{Hedgehog_MFtheory_Binz_prb06, Hedgehog_MFtheory_Park_11, Hedgehog_MCtheory_Yang_16, MnGe_Kanazawa_16, Hedgehog_MCtheory_Okumura_19, SFO_Ishiwata_20, hedgehog_AK_prb_21}. In this paper, we theoretically investigate the stability of the hedgehog lattice recently shown to emerge in breathing pyrochlore antiferromagnets \cite{hedgehog_AK_prb_21} against the inclusion of ferromagnetic exchange interactions, and discuss a magnetic-field-induced total chirality associated with the anomalous Hall effect of chirality origin.

The magnetic hedgehog is a spin texture characterized by an integer topological charge which corresponds to, in units of $4\pi$, the total solid angle subtended by all the spins involved. As the hedgehog has a singular point at its texture center, it is sometimes called the magnetic monopole. The hedgehog lattice is an alternating periodic array of the hedgehogs (monopoles) and anti-hedgehogs (anti-monopoles) each having positive or negative nonzero topological charge, so that the net topological charge summed over the whole system is zero. Noting that the solid angle $\Omega_{ijk}$ for three spins ${\bf S}_i$, ${\bf S}_j$, and ${\bf S}_k$ is related with the scalar spin chirality $\chi_{ijk}={\bf S}_i \cdot ({\bf S}_j \times {\bf S}_k)$ via $\Omega_{ijk}=2\tan^{-1}\big[\frac{\chi_{ijk}}{1+{\bf S}_i\cdot{\bf S}_j +{\bf S}_i\cdot{\bf S}_k+{\bf S}_j\cdot{\bf S}_k} \big]$ \cite{SolidAngle_OOsterom_83}, the hedgehog lattice can be understood as a long-range order (LRO) of the scalar chirality $\chi_{ijk}$. Such a complicated noncoplanar spin structure is usually described by more than one ordering-wave-vectors, i.e., it is a multiple-${\bf Q}$ state.

A prominent aspect of the hedgehog lattice is the magnetic-field-induced anomalous Hall effect of chirality origin, the so-called topological Hall effect \cite{Hedgehog_MFtheory_Park_11, Hedgehog_transport_Zhang_16, MnGe_Kanazawa_11, MnGe_Kanazawa_16, Hedgehog_MCtheory_Okumura_19, SFO_Ishiwata_11, SFO_Ishiwata_20, hedgehog_AK_prb_21}. In contrast to the topological Hall signal for the skyrmion lattice showing a distinct nonzero value, the corresponding signal for the hedgehog lattice is zero at zero field and gradually increases with increasing field. The former signal is directly connected to the total skyrmion numbers, whereas the latter is not directly associated with the monopole/antimonopole numbers themselves, but rather with the total chirality or the total number of skyrmions generated in an intermediate spatial region between the monopoles and antimonopoles \cite{Hedgehog_MFtheory_Park_11, Hedgehog_transport_Zhang_16, Hedgehog_MCtheory_Okumura_19, hedgehog_AK_prb_21}.

The hedgehog lattice having such a field-induced total chirality is known to be stabilized in the presence of the DM interaction \cite{Hedgehog_MFtheory_Binz_prb06, Hedgehog_MFtheory_Park_11, Hedgehog_MCtheory_Yang_16, Hedgehog_MCtheory_Okumura_19}. In the non-centrosymmetric magnets having the DM interaction MnSi$_{1-x}$Ge$_x$ with $0.25<x$, the occurrence of the hedgehog lattice has been evidenced by the topological Hall effect together with the observation of multiple-${\bf Q}$ Bragg reflections \cite{MnGe_Kanazawa_11, MnGe_Kanazawa_12, MnGe_Shiomi_13, MnGe_Tanigaki_15, MnGe_Kanazawa_16, MnGe_Kanazawa_17, MnSiGe_Fujishiro_19}. Recently, the hedgehog lattice is found also in the DM-free centrosymmetric magnet SrFeO$_3$ \cite{SFO_Ishiwata_11,SFO_Ishiwata_20}, but the ordering mechanism is still unclear and no other candidate compounds have been reported so far. In view of such a situation, we have searched for a new mechanism other than the DM interaction. Previously, we theoretically demonstrated that the hedgehog lattice is realized in breathing-pyrochlore antiferromagnets in the absence of the DM interaction \cite{hedgehog_AK_prb_21}. In this work, we extend our previous theoretical work, examining the stability of the hedgehog lattice in a wider parameter space.

The breathing pyrochlore lattice is a three-dimensional network consisting of an alternation array of corner-sharing small and large tetrahedra \cite{BrPyro_Okamoto_13, BrPyro_Tanaka_14, BrPyro_Nilsen_15, BrPyro_Saha_16, BrPyro_Lee_16, BrPyro_Saha_17, qBrPyro_Kimura_14}. The characteristic feature of this breathing lattice is that the nearest neighbor (NN) exchange interactions on small and large tetrahedra $J_1$ and $J_1'$ take different values due to the bond-length difference. In our previous work, we showed that a quadruple-${\bf Q}$ state characterized by the four ordering vectors of ${\bf Q}_1=(\frac{1}{2}, \frac{1}{2}, \frac{1}{2}), \, {\bf Q}_2=(-\frac{1}{2}, \frac{1}{2}, \frac{1}{2}), \, {\bf Q}_3=(\frac{1}{2}, -\frac{1}{2}, \frac{1}{2})$, and ${\bf Q}_4=(\frac{1}{2}, \frac{1}{2}, -\frac{1}{2})$ in units of $\frac{2\pi}{a}$ in the basis of the cubic unit cell with side length $a$, which is realized for a large third NN antiferromagnetic (AFM) interaction along the bond direction $J_3$ \cite{Site_AK_16, Site_AK_19}, becomes the hedgehog-lattice state in the breathing case of $J_1'/J_1 <1$, while in the uniform case of $J_1'/J_1 =1$, it is a collinear state favored by thermal fluctuations, where both $J_1$ and $J_1'$ are assumed to be AFM \cite{hedgehog_AK_prb_21}. 
On the other hand, among so-far reported breathing-pyrochlore magnets, chromium sulfides possess ferromagnetic (FM) $J_1$ and/or $J_1'$ \cite{BrPyro_Sulfides_Okamoto_18, BrPyro_Sulfides_Pokharel_18, FirstPrinciple_Ghosh_npj_19}: FM $J_1$ and FM $J_1'$ for Li(Ga,In)Cr$_4$S$_8$, and AFM $J_1$ and FM $J_1'$ for CuInCr$_4$S$_8$. Then, the naive question is whether the hedgehog lattice is robust against the inclusion of FM $J_1$ and/or $J_1'$. In addition, effects of a magnetic field in such a situation would also be an interesting issue. 

\begin{figure}[t]
\begin{center}
\includegraphics[scale=0.52]{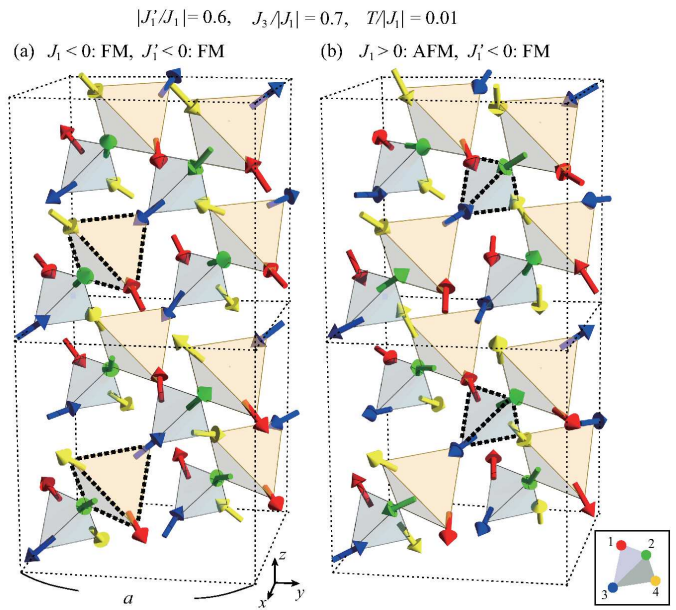}
\caption{Monte Carlo snapshots of quadruple-${\bf Q}$ $(\frac{1}{2},\frac{1}{2},\frac{1}{2})$ hedgehog-lattice spin textures on the breathing pyrochlore lattice, where the spin configurations are obtained at $H=0$ and $T/|J_1|=0.01$ for $|J'_1/J_1|=0.6$ and $J_3/|J_1|=0.7$ with (a) FM $J_1$ and FM $J_1'$ and (b) AFM $J_1$ and FM $J_1'$. Red, green, blue, and yellow arrows represent spins on the corners 1, 2, 3, and 4 of the small tetrahedra shown in the inset, respectively. The lower and upper tetrahedra outlined black dots correspond to the hedgehog (monopole) and the anti-hedgehog (anti-monopole), respectively. \label{fig:hedgehog_snapshot}}
\end{center}
\end{figure}

\begin{figure*}[t]
\begin{center}
\includegraphics[scale=0.34]{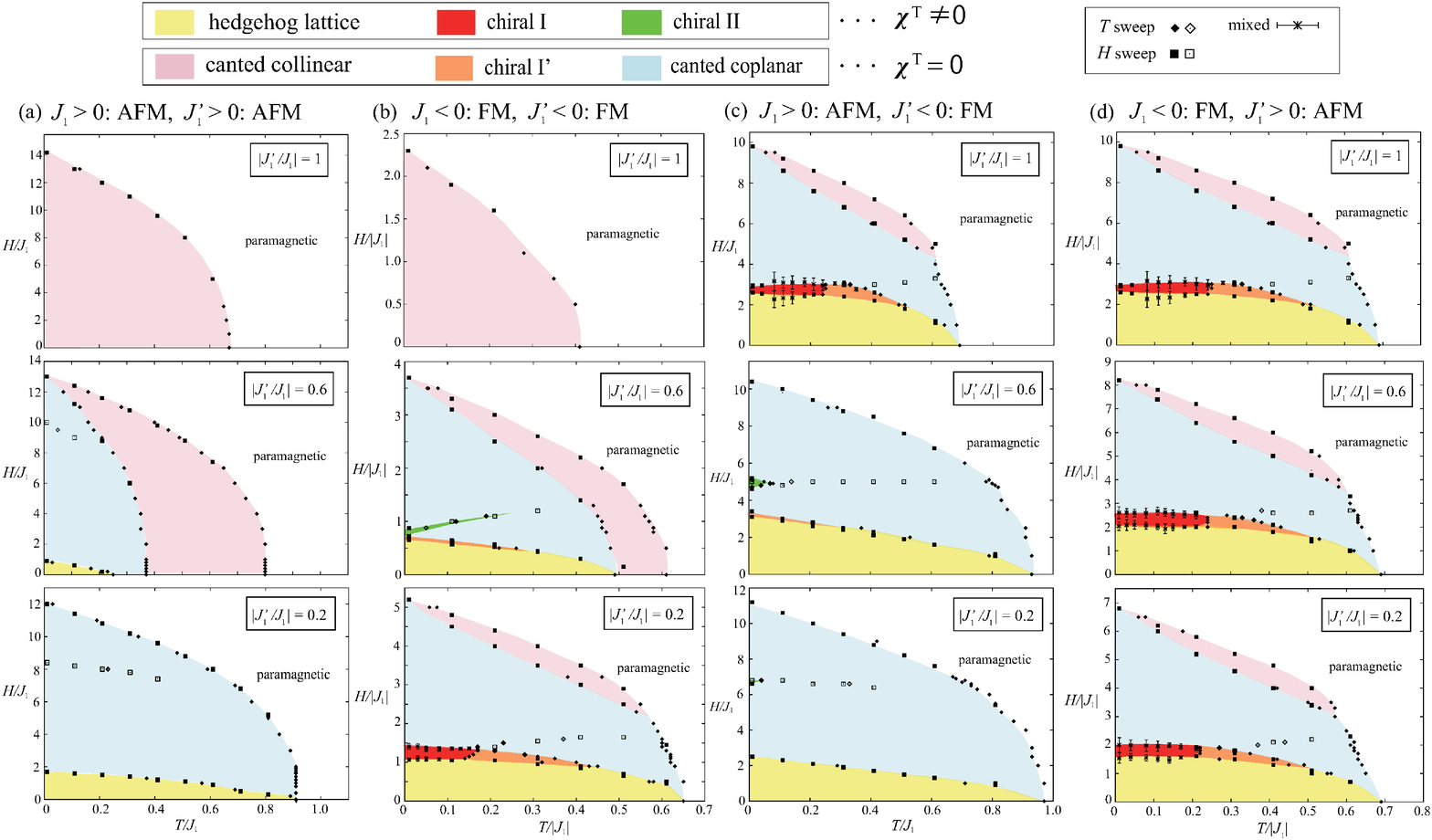}
\caption{Temperature and magnetic-field phase diagrams obtained in the cases of (a) AFM $J_1>0$ and AFM $J_1'>0$, (b) FM $J_1<0$ and FM $J_1'<0$, (c) AFM $J_1>0$ and FM $J_1'<0$, and (d) FM $J_1<0$ and AFM $J_1'>0$ for $|J_1'/J_1|=1$ (top), $|J_1'/J_1|=0.6$ (middle), and $|J_1'/J_1|=0.2$ (bottom), where the $J_3$ value is fixed to be $J_3/|J_1|=0.7$. The middle and bottom panels of (a) are taken from Ref. \cite{hedgehog_AK_prb_21}. For $|J_1'/J_1|=1$, the AFM $J_1>0$ and FM $J_1'<0$ case is essentially the same as the FM $J_1<0$ and AFM $J_1'>0$ case where the roles of small and large tetrahedra are merely interchanged, so that the top panels of (c) and (d) are exactly the same except the temperature range. In the in-field hedgehog-lattice (yellow), chiral I (red), and chiral II (green) phases, the field-induced total chirality vector $\mbox{\boldmath $\chi$}^{\rm T}$ corresponding to the emergent fictitious field is nonzero, whereas in the canted collinear (pink), canted coplanar (light blue), and chiral I' (orange) phases, $\mbox{\boldmath $\chi$}^{\rm T}$ is zero. Although the canted coplanar phases can be categorized into two, i.e., the low-field and high-field ones, their ordering properties are essentially the same (for details, see Secs. IV and V). \label{fig:HT_phaseall}}
\end{center}
\end{figure*}

In this work, we show by means of Monte Carlo (MC) simulations that the quadruple-${\bf Q}$ $(\frac{1}{2},\frac{1}{2},\frac{1}{2})$ hedgehog-lattice state is stable even in the presence of FM $J_1$ and/or $J_1'$. As one can see from the MC snapshots shown in Fig. \ref{fig:hedgehog_snapshot}, the monopoles and antimonopoles are formed on large (small) tetrahedra for FM (AFM) $J_1$. In a magnetic field, there appear six quadruple-${\bf Q}$ $(\frac{1}{2},\frac{1}{2},\frac{1}{2})$ states: canted collinear, canted coplanar, hedgehog-lattice, chiral I, chiral I', and chiral II phases (see Fig. \ref{fig:HT_phaseall}) among which the hedgehog-lattice, chiral I, and chiral II phases exhibit nonzero total chirality $\mbox{\boldmath $\chi$}^{\rm T}$ corresponding to the emergent fictitious magnetic field. Interestingly, in contrast to the hedgehog-lattice, the chiral I and II phases do not possess topological spin textures in spite of the fact that $\mbox{\boldmath $\chi$}^{\rm T}\neq 0$, which is reflected in the difference in the manner how $\mbox{\boldmath $\chi$}^{\rm T}$ is induced by the field.   

The outline of this paper is as follows: In Sec. II, we introduce the model and relevant physical quantities. This is followed by Sec. III in which the stability of the hedgehog lattice at zero field is discussed based on MC and mean-field results. Concerning the effects of an applied magnetic field, we first give a brief summary of in-field phases in Sec. IV, and then, discuss the detailed spin and chirality structures in the in-field phases in Sec. V and VI, respectively. We end the paper with summary and discussion in Sec. VII. Supplementary informations of the chirality in a magnetic field are provided in Appendices A, B, and C.

\section{Model and relevant physical quantities}
In this work, we consider the $J_1$-$J_3$ classical Heisenberg model on the breathing pyrochlore lattice which is given by
\begin{eqnarray}\label{eq:Hamiltonian}
{\cal H} &=& J_1 \sum_{\langle i,j \rangle_S} {\bf S}_i \cdot {\bf S}_j + J_1' \sum_{\langle i,j \rangle_L} {\bf S}_i \cdot {\bf S}_j \nonumber\\
&& + J_{3} \sum_{\langle \langle i,j \rangle \rangle} {\bf S}_i \cdot {\bf S}_j + H\sum_i S_i^z,
\end{eqnarray}  
where $\langle \rangle_{S(L)}$ and $\langle \langle  i,j \rangle \rangle$ denote the summations over site pairs on the small (large) tetrahedra and the third NN pairs along the bond direction, respectively, and $H$ is a magnetic field applied in the $z$-direction in the spin space. In the model, the breathing bond-alternation of the lattice is characterized by $J_1'/J_1$, and the third NN AFM interaction along the bond direction $J_3$ is essential for the occurrence of a quadruple-${\bf Q}$ $(\frac{1}{2},\frac{1}{2},\frac{1}{2})$ state characterized by ${\bf Q}_1$, ${\bf Q}_2$, ${\bf Q}_3$, and ${\bf Q}_4$. Since at least in the case of AFM $J_1>0$ and $J_1'>0$, the additional second NN interaction $J_2$ is irrelevant to the occurrence of the hedgehog lattice \cite{hedgehog_AK_prb_21}, it is not incorporated in the spin Hamiltonian (\ref{eq:Hamiltonian}).

In order to examine the stability of the hedgehog lattice against the sign changes in $J_1$ and $J_1'$, we use the same $J_3$ value of $J_3/|J_1|=0.7$ as in the AFM $J_1$ and $J_1'$ case discussed in the previous paper, and change the signs of $J_1$ and $J_1'$ for the three fixed values of $|J_1'/J_1|=1$, 0.6, and 0.2. Note that $J_1'/J_1=1$ corresponds to the uniform pyrochlore lattice, while $J_1'/J_1=-1$ does not; all parameter sets except $J_1'/J_1=1$ correspond to the breathing pyrochlore lattice. It should also be noted that for $J_1'/J_1=-1$, the $J_1>0$ and $J_1'<0$ case is essentially the same as the $J_1<0$ and $J_1'>0$ case where the roles of small and large tetrahedra are merely interchanged. 

To investigate the spin ordering, we first introduce the spin structure factor $F_{S}({\bf q})=F_{S\parallel}({\bf q})+F_{S\perp}({\bf q})$ defined by 
\begin{eqnarray}\label{eq:F_S}
F_{S\parallel}({\bf q}) &=& \Big\langle \Big| \frac{1}{N} \sum_i  S^z_i \, e^{i{\bf q}\cdot{\bf r}^0_i}\Big|^2\Big\rangle, \nonumber\\
F_{S\perp}({\bf q}) &=& \Big\langle \sum_{\nu=x,y} \Big| \frac{1}{N} \sum_i  S^\nu_i \, e^{i{\bf q}\cdot{\bf r}^0_i}\Big|^2\Big\rangle, 
\end{eqnarray}
where $N$ is a total number of spins which is related to the linear system size $L$ via $N=16L^3$ as the cubic unit cell contains 16 sites, and $\langle {\cal O} \rangle$ denotes the thermal average of a physical quantity ${\cal O}$. Noting that the magnetic field $H$ is applied in the $S^z$ direction, we have introduced $F_{S\parallel}({\bf q})$ for the $S^z$ component and $F_{S\perp}({\bf q})$ for the $S^xS^y$-plane component. Since the breathing bond-alternation has already been incorporated in the spin Hamiltonian (\ref{eq:Hamiltonian}) in the form of the nonequivalent $J_1$ and $J_1'$, we have taken ${\bf r}^0_i$ in Eq. (\ref{eq:F_S}) as a site position of the {\it uniform} pyrochlore lattice ignoring the bond-length alternation for simplicity. As the chirality sector is also important, we introduce the chirality structure factor defined by
\begin{equation}\label{eq:F_C}
F_{C}({\bf q}) = \big\langle \big| \frac{1}{N/2} \sum_l  \chi({\bf R}_l) \, e^{i{\bf q}\cdot{\bf R}_l}\big|^2\big\rangle,
\end{equation}
where the summation is taken over all the small and large tetrahedra. The scalar chirality of the $l$th tetrahedron with its center-of-mass position ${\bf R}_l$ is defined by $\chi({\bf R}_l) = \sum_{i,j,k \in l \, {\rm th} \, {\rm tetra}} \chi_{ijk}$ and the order of $i$, $j$, and $k$ is defined in the anticlockwise direction with respect to the normal vector of the triangle formed by the three sites $i$, $j$, and $k$, $\hat{n}_{ijk}$, pointing outward from ${\bf R}_l$. In the same manner, we define the solid angle subtended by the four spins on the tetrahedron as $\Omega({\bf R}_l)=\sum_{i,j,k \in l \, {\rm th} \, {\rm tetra}}\Omega_{ijk}$. 

In addition to the spin and chirality structure factors, the spin collinearity $P$ and the $S^xS^y$-plane nematicity $P_{\perp 2}$ also provide useful informations of the spin structures. They are given by
\begin{eqnarray}\label{eq:OP_nematic}
P &=& \frac{3}{2} \Big\langle \frac{1}{N^2}\sum_{i,j} \big( {\bf S}_i\cdot{\bf S}_j\big)^2 - \frac{1}{3} \Big\rangle, \nonumber\\
P_{\perp 2} &=& \frac{3}{4}\Big\langle \Big(\frac{1}{N}\sum_i Q_i^{x^2-y^2}\Big)^2+\Big(\frac{1}{N}\sum_i Q_i^{xy}\Big)^2\Big\rangle, \nonumber\\
Q_i^{x^2-y^2} &=& (S_i^x)^2-(S_i^y)^2, \qquad Q_i^{xy}=2S_i^xS_i^y .
\end{eqnarray}
The nematic order parameter $P_{\perp,2}$ measures the 2-fold breaking of rotational symmetry in the $S^xS^y$ plane perpendicular to the magnetic field, so that it takes a nonzero value when there exists a characteristic axis within the $S^xS^y$ spin plane \cite{Bond_Shannon_10, Site_AGK_21}.  

In general, when localized spins ${\bf S}_i$ are weakly coupled to conduction electrons in a metallic system, an anomalous Hall effect is caused by the total chirality {\it vector} $\mbox{\boldmath $\chi$}^{\rm T}$ summed over the whole system which is given by
\begin{equation}\label{eq:total_chi}
\mbox{\boldmath $\chi$}^{\rm T} =\Big\langle \sum_{\langle i,j,k \rangle_{S,L}} \chi_{ijk} \, \hat{n}_{ijk} \Big\rangle.
\end{equation}
Here, the chirality-driven anomalous Hall conductivity $\sigma^{\rm T}_{\mu\nu}$ is related to $\mbox{\boldmath $\chi$}^{\rm T}$ via $\sigma^{\rm T}_{\mu\nu} \propto \epsilon_{\mu\nu\rho} \chi^{\rm T}_\rho$ \cite{THE_Tatara_02}, so that the total chirality vector $\mbox{\boldmath $\chi$}^{\rm T}$ corresponds to the emergent fictitious field. Although according to Ref. \cite{THE_Tatara_02}, $\mbox{\boldmath $\chi$}^{\rm T}$ should be calculated by taking further NN sites into account, we have restricted only to the NN triads because the dominant contribution comes from such short-distance triads. As one can see from Eq. (\ref{eq:total_chi}), the anomalous Hall effect of chirality origin {\it in three dimensions} is {\it not directly} connected to the scalar spin chirality $\chi_{ijk}$ itself but rather the spin chirality multiplied by a geometrical factor $\hat{n}_{ijk}$, which is in sharp contrast to two-dimensional systems where $\hat{n}_{ijk}$ is always perpendicular to the two-dimensional lattice plane and thereby, $\mbox{\boldmath $\chi$}^{\rm T}$ is also perpendicular to the lattice plane with its strength being directly related to the spin chirality. We also note that $F_{C}({\bf q})$ in Eq. (\ref{eq:F_C}) does not involve the geometrical factor $\hat{n}_{ijk}$ and thus, is not directly associated with $\mbox{\boldmath $\chi$}^{\rm T}$.

To calculate the physical quantities introduced above, we perform MC simulations, where 2$\times 10^5$ sweeps are carried out under the periodic boundary condition and the first half is discarded for thermalization. Our 1 MC sweep consists of 1 heatbath sweep and successive 10 overrelaxation sweeps, and observations are done at every MC sweep. The statistical average is taken over 4 independent runs starting from different random initial configurations. By carefully analyzing the above quantities and fundamental ones such as the specific heat $C = \frac{1}{T^2N}\big( \langle {\cal H}^2\rangle - \langle {\cal H} \rangle^2\big)$ and the magnetization $m = \langle | \frac{1}{N}\sum_{i} {\bf S}_i | \rangle$, we identify low-temperature ordered phases. 

In obtaining the temperature and magnetic field phase diagrams shown in Fig. \ref{fig:HT_phaseall}, we also use the mixed phase method \cite{MixedMethod_Creutz_79, Site_AGK_21} to determine the low-temperature phase boundaries between the chiral I phase and the lower-field hedgehog-lattice and the higher-field coplanar phases, since a first-order character of the transition and the associated hysteresis is relatively strong in this regime. In the quite low-temperature regime for $J_1'/J_1=-1$ where even the mixed phase method does not work, we determine the phase boundary by comparing the energies of the competing phases.
We also note that in calculating the monopole density, we evaluate the solid angle for each tetrahedron $\Omega({\bf R}_l)$ by performing the short-time average over 200 MC steps to reduce the thermal noise, where we first evaluate $\Omega({\bf R}_l)$ by using spin configurations averaged over 10 MC sweeps and then take an average over 20 samples of the so-obtained solid angle to avoid counting monopoles accidentally generated by thermal fluctuations.

\section{Stability of the hedgehog lattice at $H=0$}
In the previous work, we showed that in the case of AFM $J_1$ and $J_1'$, three types of quadruple-${\bf Q}$ $(\frac{1}{2},\frac{1}{2},\frac{1}{2})$ states, collinear, coplanar, and noncoplanar hedgehog-lattice states, appear depending on the value of $J_1'/J_1$. In the uniform case of $J_1'/J_1=1$, only the collinear phase is realized, whereas in the breathing case of $J_1'/J_1<1$, the hedgehog-lattice phase is realized at the lowest temperature and the coplanar phase can appear at higher temperatures \cite{hedgehog_AK_prb_21}. 
In this section, we will discuss the stability of the hedgehog lattice at $H=0$ against the sign changes in $J_1$ and $J_1'$. As we will see below, MC results show that the hedgehog lattice can be realized even for FM $J_1$ and/or $J_1'$, which can readily be understood from a mean-field result. 

\subsection{Monte Carlo result}
\begin{figure}[t]
\begin{center}
\includegraphics[scale=0.62]{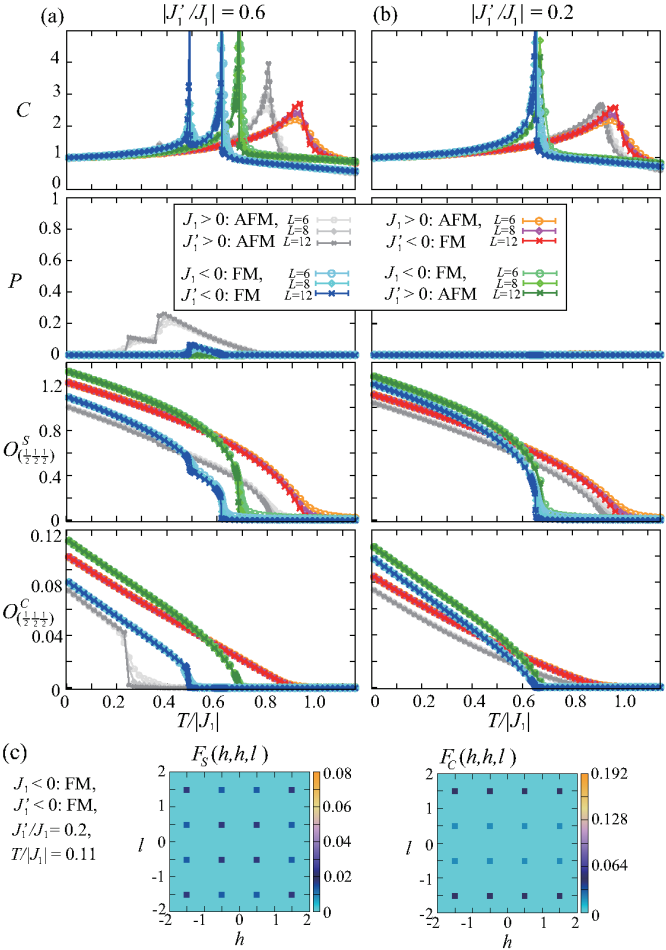}
\caption{MC results obtained at $H=0$ for $J_3/|J_1|=0.7$. (a) and (b) The temperature dependence of the specific heat $C$ (the first panel from the top), the spin collinearity $P$ (the second one), and the average spin and chirality Bragg intensities $O^S_{(\frac{1}{2}\frac{1}{2}\frac{1}{2})}$ and $O^C_{(\frac{1}{2}\frac{1}{2}\frac{1}{2})}$ (the third and fourth ones) for (a) $|J_1'/J_1|=0.6$ and (b) $|J_1'/J_1|=0.2$. Grayish, reddish, bluish, and greenish colored symbols denote data for AFM $J_1$ and AFM $J_1'$, AFM $J_1$ and FM $J_1'$, FM $J_1$ and FM $J_1'$, and FM $J_1$ and AFM $J_1'$, respectively. The grayish data are taken from Ref. \cite{hedgehog_AK_prb_21}. (c) Spin and chirality structure factors $F_S({\bf q})$ (left) and $F_C({\bf q})$ (right) in the $(h,h,l)$ plane obtained at $T/|J_1|=0.11$ for FM $J_1$, FM $J_1'$, $J_1'/J_1=0.2$, and $L=12$. \label{fig:H0Tdep}}
\end{center}
\end{figure}
Figure \ref{fig:H0Tdep} shows MC results obtained in the breathing cases of $|J_1'/J_1|=0.6$ and $0.2$ for the four different combinations, AFM $J_1$ and AFM $J_1'$ (grayish colored symbols), AFM $J_1$ and FM $J_1'$ (reddish ones), FM $J_1$ and FM $J_1'$ (bluish ones), and FM $J_1$ and AFM $J_1'$ (greenish ones). Since in the uniform case of $J_1'/J_1=1$ with FM $J_1<0$ and $J_1'<0$, merely the collinear state appears as in the AFM uniform case of $J_1=J_1'>0$, only the breathing cases are shown in Fig. \ref{fig:H0Tdep}. As one can see from the top panels in Figs. \ref{fig:H0Tdep} (a) and (b), the transition temperature indicated by the specific-heat sharp peak is higher in the AFM $J_1$ cases than in the FM $J_1$ cases, and is slightly enhanced by the mixing of AFM and FM interactions, i.e., $J_1$ and $J_1'$ of opposite signs. One can see from the second panels from the top in Figs. \ref{fig:H0Tdep} (a) and (b) that in all the cases, a non-collinear state with $P=0$ is realized in the lower-temperature ordered phase. In the $|J_1'/J_1|=0.6$ case with FM $J_1<0$ and $J_1'<0$ [see the bluish symbols in Fig. \ref{fig:H0Tdep} (a)], the collinear state with $P \neq 0$ appears just below the transition from the paramagnetic phase, whereas in the $|J_1'/J_1|=0.6$ case with AFM $J_1>0$ and $J_1'>0$ [see the grayish symbols in Fig. \ref{fig:H0Tdep} (a)], a coplanar state is realized in the intermediate temperature window between the higher-temperature collinear and lower-temperature $P=0$ states \cite{hedgehog_AK_prb_21}.

As exemplified by Fig. \ref{fig:H0Tdep} (c), the spin structure factors $F_S({\bf q})$ in all the low-temperatures phases exhibit multiple Bragg peaks at ${\bf Q}_1=(\frac{1}{2}, \frac{1}{2}, \frac{1}{2}), \, {\bf Q}_2=(-\frac{1}{2}, \frac{1}{2}, \frac{1}{2}), \, {\bf Q}_3=(\frac{1}{2}, -\frac{1}{2}, \frac{1}{2})$, and ${\bf Q}_4=(\frac{1}{2}, \frac{1}{2}, -\frac{1}{2})$, suggesting that they are quadruple-${\bf Q}$ $(\frac{1}{2}, \frac{1}{2}, \frac{1}{2})$ states. Actually, one can see from the third panels from the top in Figs. \ref{fig:H0Tdep} (a) and (b) that the normalized Bragg intensity averaged over the four ordering vectors $O^S_{(\frac{1}{2}\frac{1}{2}\frac{1}{2})} = 16 \sum_{h,k,l=\pm 1/2} F_S (h,k,l)/8$ develops on entering the low-temperature ordered phase from the high-temperature paramagnetic phase. Furthermore, in the $P=0$ state, the chirality structure factor $F_C({\bf q})$ exhibits Bragg peaks at $(\pm\frac{1}{2}, \pm\frac{1}{2}, \pm\frac{1}{2})$ [see the right panel of Fig. \ref{fig:H0Tdep} (c)]. As one can see from the bottom panels in Figs. \ref{fig:H0Tdep} (a) and (b), the averaged Bragg intensity in the chirality sector $O^C_{(\frac{1}{2}\frac{1}{2}\frac{1}{2})} = \frac{1}{4}\sum_{h,k,l=\pm 1/2} F_C (h,k,l)$ is nonzero only in the $P=0$ state, indicating that a chirality order with a noncoplanar spin structure is realized in this phase. As will be explained below, this noncoplanar structure turns out to be the hedgehog lattice.

\begin{figure}[t]
\begin{center}
\includegraphics[width=\columnwidth]{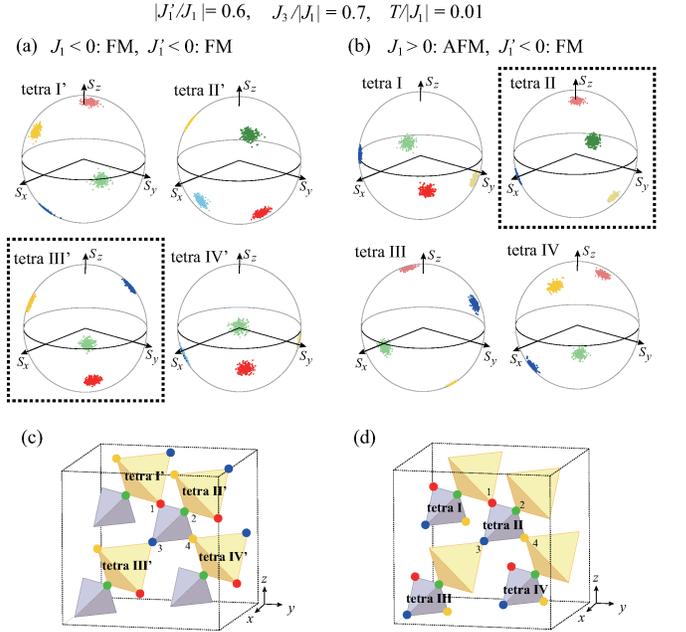}
\caption{MC snapshots of spins mapped onto a unit sphere. (a) and (b) correspond to Figs. \ref{fig:hedgehog_snapshot} (a) and (b), respectively, where the whole system spins belonging to the lower-type cubic unit cell in Fig. \ref{fig:hedgehog_snapshot} are shown. In (a), tetra's I', II', III', and IV' represent the four large tetrahedra in the cubic unit cell shown in (c), whereas in (b), tetra's I, II, III, and IV represent the four small ones shown in (d). The color notations are the same as those in Fig. \ref{fig:hedgehog_snapshot}: red, green, blue, and yellow arrows represent
spins on the four sublattices 1, 2, 3, and 4 shown in (c) and (d). In (a) and (b), the tetrahedra enclosed by black dots correspond to the magnetic hedgehog (monopole). \label{fig:hedgehog_snapmap}}
\end{center}
\end{figure}

Figures \ref{fig:hedgehog_snapshot} (a) and (b) show MC spin snapshots obtained in the noncoplanar phases for $|J_1'/J_1|=0.6$ with FM $J_1$ and $J_1'$, and AFM $J_1$ and FM $J_1'$, respectively. One can see from Fig. \ref{fig:hedgehog_snapshot} (a) that all-in and all-out spin configurations are realized on the large tetrahedra enclosed by black dots in the upper and lower cubic unit cells, respectively. The total solid angle subtended by the four spins on the all-out (all-in) tetrahedron is $4\pi$ ($-4\pi$), so that the all-out (all-in) tetrahedron corresponds to the monopole (antimonopole). Since the quadruple-${\bf Q}$ $(\frac{1}{2}, \frac{1}{2}, \frac{1}{2})$ state is a 32-sublattice state consisting of the alternating array of the two cubic unit cells, i.e., the upper and lower cubic unit cells in Fig. \ref{fig:hedgehog_snapshot}, it is definitely the hedgehog lattice consisting of the alternating array of the monopoles and antimonopoles. As the magnetic unit cell contains 16 tetrahedra, the monople (or equivalently, antimonopole) density $n_+(=n_-)$ is $\frac{1}{16}$. To identify the monopole tetrahedra, it is convenient to map spins onto a unit sphere. Figure \ref{fig:hedgehog_snapmap} shows the mapped whole-system spins belonging to the lower-type cubic unit cell in Fig. \ref{fig:hedgehog_snapshot}. Note that spins belonging to the upper-type cubic unit cell are obtained by merely replacing ${\bf S}_i$ with $-{\bf S}_i$. As readily seen from Fig. \ref{fig:hedgehog_snapmap} (a), spins span the total solid angle of $4\pi$ on tetra III', while not on the remaining three. In the AFM $J_1$ case shown in Figs. \ref{fig:hedgehog_snapshot} (b) and \ref{fig:hedgehog_snapmap} (b), the monopoles and antimonoples are formed on the small tetrahedra. 

As is well known, the AFM NN interaction tends to orient four spins on a tetrahedron ${\bf S}_1$, ${\bf S}_2$, ${\bf S}_3$, and ${\bf S}_4$ such that the constraint ${\bf S}_1+{\bf S}_2+{\bf S}_3+{\bf S}_4=0$ be satisfied. Since the all-in and all-out configurations satisfy this constraint, the monopole and antimonopole tend to stay at the tetrahedra with stronger AFM $J_1$ or $J_1'$, or equivalently, weaker FM $J_1$ or $J_1'$. Thus, in the FM and AFM $J_1$ cases with $|J_1|>|J_1'|$, the monopoles are formed on the large and small tetrahedra, respectively. 

In Fig. \ref{fig:hedgehog_snapshot}, one notices that spins belonging to each of the four sublattices corresponding to the four corners of the tetrahedron constitute almost up-down-up-down chains running along all the bond directions (see the same color arrows in Fig. \ref{fig:hedgehog_snapshot}). Suppose that the spin polarization vector of the up-down-up-down chains on sublattice $\mu$ be $\hat{P}_\mu$. In the hedgehog-lattice phase, $\hat{P}_1$, $\hat{P}_2$, $\hat{P}_3$, and $\hat{P}_4$ orient in the different directions, resulting in the in- and out-type spin configuration. It should be emphasized here that in the present system, the isotropic Heisenberg spins spontaneously form the spin-ice-type noncoplanar spin structure, which is in sharp contrast to the spin-ice system where the spin orientation is restricted to the in- and out-directions from the beginning due to the local magnetic anisotropy. Actually, reflecting the Heisenberg nature, the collinear and coplanar states are also possible depending on the value of $J_1'/J_1$. All the three states, hedgehog-lattice, coplanar, and collinear states, are quadruple-${\bf Q}$ states described as superpositions of the four up-down-up-down chains. The difference among the three consists in the relative angles among $\hat{P}_1$, $\hat{P}_2$, $\hat{P}_3$, and $\hat{P}_4$. In the next subsection, we will discuss how the breathing bond-alternation, i.e., the nonequivalence of $J_1$ and $J_1'$, favors the noncoplanar superposition pattern, based on the mean-field result.

\subsection{Mean field analysis}
For AFM $J_1>0$ and $J_1'>0$, we have already derived the Ginzburg-Landau (GL) free energy for the $(\frac{1}{2}\frac{1}{2}\frac{1}{2})$ state in Ref. \cite{hedgehog_AK_prb_21}. Since the derivation with the use of the mean-field approximation is valid irrespective of the signs of $J_1$ and $J_1'$, the GL free energy for FM $J_1$ and/or $J_1'$ takes the same form as that for AFM $J_1$ and $J_1'$. Thus, as discussed in  Ref. \cite{hedgehog_AK_prb_21}, the relative angles among $\hat{P}_1$, $\hat{P}_2$, $\hat{P}_3$, and $\hat{P}_4$ are determined by a GL quartic term $\delta f_4$ given by
\begin{eqnarray}\label{eq:GL}
\delta f_4 &=& \frac{9T}{640}{\overline S}^4 \bigg[ A_2 \Big\{ 3 + ({\hat P}_1\cdot{\hat P}_2)^2 + ({\hat P}_1\cdot{\hat P}_3)^2  \nonumber\\
&& + ({\hat P}_1\cdot{\hat P}_4)^2 + ({\hat P}_2\cdot{\hat P}_3)^2 + ({\hat P}_2\cdot{\hat P}_4)^2 + ({\hat P}_3\cdot{\hat P}_4)^2 \Big\} \nonumber\\
&-& 2A_3 \Big\{ ({\hat P}_1\cdot{\hat P}_2)({\hat P}_3\cdot{\hat P}_4) + ({\hat P}_1\cdot{\hat P}_3)({\hat P}_2\cdot{\hat P}_4) \nonumber\\
&& + ({\hat P}_1\cdot{\hat P}_4)({\hat P}_2\cdot{\hat P}_3) \Big\}  \bigg]. 
\end{eqnarray}
Here, ${\overline S}$ corresponds to the thermal-averaged spin length and the coefficients are given by $A_2 = \frac{16\varepsilon^2(1+\varepsilon^2)}{(1+3\varepsilon^2)^2}$ and $A_3 = \frac{16\varepsilon^3}{(1+3\varepsilon^2)^2}$
with 
\begin{equation}\label{eq:varepsilon}
\varepsilon = \frac{J_1+J'_1+4J_3}{3(J_1-J'_1)}\bigg[ \sqrt{1 + 3 \Big( \frac{J_1-J'_1}{J_1+J'_1+4J_3}\Big)^2 } -1 \bigg].
\end{equation}
In the uniform case of $J_1=J_1'$, the $\delta f_4$ term vanishes because $\varepsilon=0$ and thereby, $A_2=A_3=0$, so that the relative angles among $\hat{P}_1$, $\hat{P}_2$, $\hat{P}_3$, and $\hat{P}_4$ cannot be determined and all the superposition patterns are energetically degenerate.  
Such a degeneracy is lifted in the breathing case of $J_1 \neq J_1'$ where $\delta f_4$ is active because $\varepsilon$ becomes nonzero. The minimization condition for $\delta f_4$ yields ${\hat P}_\mu \cdot {\hat P}_\nu =-1/3$ ($\mu \neq \nu$), pointing to the emergence of the noncoplanar hedgehog-lattice state. As $|\varepsilon|$ increases, the coefficients $A_2$ and $A_3$ take larger values and correspondingly, the contribution from $\delta f_4$ becomes larger, stabilizing the hedgehog-lattice state more firmly. Thus, the stability of the hedgehog-lattice phase is governed by the dimensionless parameter $\varepsilon$, whereas the breathing bond-alternation itself is characterized by the ratio $J_1'/J_1$. 

To see how $J_1$ and $J_1'$ are related with $\varepsilon$, it is convenient to consider the limiting case of $|J_1-J_1'| \ll J_1+J_1'+4J_3$ where $\varepsilon$ can be approximated as
\begin{equation}\label{eq:varepsilon_app}
\varepsilon \simeq \frac{1}{2} \frac{J_1-J'_1}{J_1+J'_1+4J_3}.
\end{equation}
One can see from Eq. (\ref{eq:varepsilon_app}) that for a fixed value of $|J_1'/J_1|$, $|\varepsilon|$ takes a larger value when $J_1$ and $J_1'$ have opposite signs because the absolute value of the numerator becomes larger. In the AFM $J_1$ case, the effect of the sign change in $J_1'$ becomes more remarkable: not only the numerator but also the denominator tends to increase $|\varepsilon|$, resulting in a higher transition temperature into the hedgehog-lattice phase, as observed in the MC result shown in Figs. \ref{fig:H0Tdep} (a) and (b).

\section{Brief summary of magnetic-field effects}
In the previous section, we show that the hedgehog lattice is stable even for FM $J_1$ and/or $J_1'$ at zero field. Below, we will discuss in-field properties of the spin Hamiltonian (\ref{eq:Hamiltonian}), putting particular emphasis on the stability of the hedgehog-lattice phase against the magnetic field and the field-induced total chirality $\mbox{\boldmath $\chi$}^{\rm T}$. This section focuses on the basic properties of various in-field phases. The detailed spin and chirality structures in these phases will be discussed in the subsequent sections. 

\begin{figure}[t]
\begin{center}
\includegraphics[width=\columnwidth]{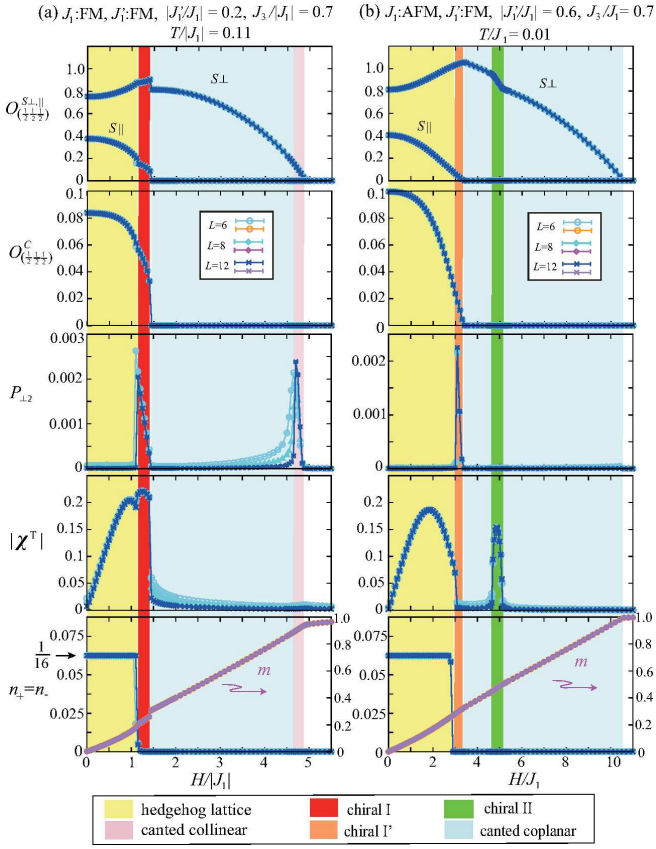}
\caption{The field dependence of various physical quantities obtained in the MC simulations at (a) $T/|J_1|=0.11$ for $|J_3/J_1|=0.7$ and $|J_1'/J_1|=0.2$ with FM $J_1$ and $J_1'$ and (b) $T/J_1=0.01$ for $J_3/J_1=0.7$ and $|J_1/J_1'|=0.6$ with AFM $J_1$ and FM $J_1'$. Top panel: the averaged Bragg intensities for the $S^xS^y$ and $S^z$ components $O^{S\perp}_{(\frac{1}{2}\frac{1}{2}\frac{1}{2})}$ and $O^{S\parallel}_{(\frac{1}{2}\frac{1}{2}\frac{1}{2})}$. The second panel from the top: the averaged chirality Bragg intensity $O^C_{(\frac{1}{2}\frac{1}{2}\frac{1}{2})}$. The third panel from the top: the $S^xS^y$-plane spin nematicity $P_{\perp 2}$. The fourth panel from the top: the total chirality associated with the Hall effect of chirality origin $|\mbox{\boldmath $\chi$}^{\rm T}|$. Bottom panel: the monopole (or equivalently, antimonopole) density $n_+(=n_-)$ and the magnetization $m$. \label{fig:Hdep}}
\end{center}
\end{figure}
\begin{figure}[t]
\begin{center}
\includegraphics[scale=0.53]{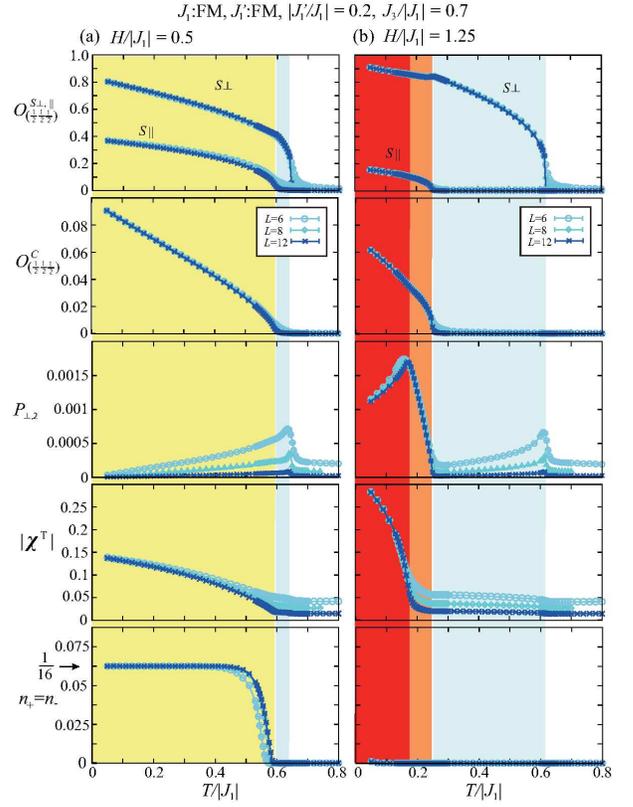}
\caption{The temperature dependence of $O^{S\perp}_{(\frac{1}{2}\frac{1}{2}\frac{1}{2})}$ and $O^{S\parallel}_{(\frac{1}{2}\frac{1}{2}\frac{1}{2})}$, $O^C_{(\frac{1}{2}\frac{1}{2}\frac{1}{2})}$, $P_{\perp 2}$, $|\mbox{\boldmath $\chi$}^{\rm T}|$, and $n_+ =n_-$ (from top to bottom) obtained in the MC simulations at (a) $H/|J_1|=0.5$ and (b) $H/|J_1|=1.25$ for $J_3/|J_1|=0.7$ and $|J_1'/J_1|=0.2$ with FM $J_1$ and $J_1'$. The color notations for the stability regions of the ordered phases are the same as those in Figs. \ref{fig:HT_phaseall} and \ref{fig:Hdep}. \label{fig:HTdep}}
\end{center}
\end{figure}

Figure \ref{fig:HT_phaseall} shows the temperature and magnetic-field phase diagrams obtained for the same parameter set as that for Fig. \ref{fig:H0Tdep}, where the results for the breathing cases of AFM $J_1$ and $J_1'$ [the middle and bottom panels of Fig. \ref{fig:HT_phaseall} (a)] are taken from Ref. \cite{hedgehog_AK_prb_21}. Note that the $J_1'/J_1=1$ cases [the top panels of Figs. \ref{fig:HT_phaseall} (a) and (b)] correspond to the uniform pyrochlore lattice, whereas the $|J_1'/J_1|=1$ cases with $J_1$ and $J_1'$ of opposite signs [the top panels of Figs. \ref{fig:HT_phaseall} (c) and (d)] correspond to the breathing pyrochlore lattice. The phase diagrams are determined from the MC results, typical examples of which are shown in Figs. \ref{fig:Hdep} and \ref{fig:HTdep}, where the field and temperature dependences of various physical quantities are presented. Figures \ref{fig:Hdep} and \ref{fig:HTdep} are associated with the bottom panel of Fig. \ref{fig:HT_phaseall} (b) and the middle panel of Fig. \ref{fig:HT_phaseall} (c).
In Fig. \ref{fig:HT_phaseall}, filled symbols denote transitions definitely identified by the physical quantities, whereas open symbols denote transition-like anomalies observed in the magnetic susceptibility for the $S^z$ component. 
 
In a magnetic field, there appear six phases: canted collinear, canted coplanar, hedgehog lattice, chiral I, chiral I', and chiral II phases whose stability regions are represented by pink, light blue, yellow, red, orange, and green in Figs. \ref{fig:HT_phaseall}, \ref{fig:Hdep}, and \ref{fig:HTdep}. All the six phases are quadruple-${\bf Q}$ $(\frac{1}{2},\frac{1}{2},\frac{1}{2})$ states with $S^xS^y$ spin components perpendicular to the field being characterized by ${\bf Q}_1$, ${\bf Q}_2$, ${\bf Q}_3$, and ${\bf Q}_4$. Indeed, the averaged intensity $O^{S\perp}_{(\frac{1}{2}\frac{1}{2}\frac{1}{2})}$ is nonzero in all the low-temperature ordered phases (see the top panels in Figs. \ref{fig:Hdep} and \ref{fig:HTdep}). In the in-field hedgehog-lattice, chiral I, and chiral I' phases, $(\frac{1}{2}, \frac{1}{2}, \frac{1}{2})$-type Bragg peaks can also be found in the $S^z$ sector and the chirality sector as well. Their averaged intensities $O^{S\parallel}_{(\frac{1}{2}\frac{1}{2}\frac{1}{2})}$ and $O^C_{(\frac{1}{2}\frac{1}{2}\frac{1}{2})}$ are, respectively, shown in the first and second panels from the top in Figs. \ref{fig:Hdep} and \ref{fig:HTdep}. 
As will be discussed in Sec. V, the spin state in the canted collinear phase is cubic-symmetric, and the spin-state symmetry in the in-field hedgehog-lattice and canted coplanar phases is reduced to tetragonal. In the chiral I and chiral I' phases, a similar tetragonal symmetry is further reduced to orthorhombic, where the rotational symmetry in the plane perpendicular to the tetragonal axis is broken. In the chiral II phase, the spin state is also orthorhombic-symmetric, but its orthorhombic nature is different from that in the chiral I and chiral I' phases. We note that the $S^xS^y$ spin nematicity $P_{\perp 2}$ detecting the existence of a characteristic axis in the $S^xS^y$ plane is nonvanishing in the canted collinear, chiral I and chiral I' phases (see the third panels from the top in Figs. \ref{fig:Hdep} and \ref{fig:HTdep}).

In the in-field hedgehog-lattice, chiral I, and chiral II phases, the total chirality $\mbox{\boldmath $\chi$}^{\rm T}$ associated with the Hall effect of chirality origin is nonzero, while in the canted collinear, canted coplanar, and chiral I' phases, $\mbox{\boldmath $\chi$}^{\rm T}$ is zero (see the fourth panels from the top in Figs. \ref{fig:Hdep} and \ref{fig:HTdep}).   
The density of the monopole tetrahedra is nonzero only in the hedgehog-lattice phase (see the bottom panels in Figs. \ref{fig:Hdep} and \ref{fig:HTdep}), although the chiral I and II phases possess the nonzero total chirality $\mbox{\boldmath $\chi$}^{\rm T} \neq 0$. 

Now, we discuss stability regions of the above in-field phases. As one can see from Fig. \ref{fig:HT_phaseall}, in the uniform case of $J_1=J_1'$, only the canted collinear phase appears. In the breathing cases of $J_1 \neq J_1'$, the hedgehog-lattice phase is realized in the low-temperature and low-field region, and the relatively wide region in the $T$-$H$ phase diagram is occupied by the canted coplanar phase. The coplanar spin structure in the higher-field side is slightly different from that in the lower-field side, although the spin structure factors for these two states do not differ qualitatively. The boundary between these two coplanar structures is indicated by open symbols in Fig. \ref{fig:HT_phaseall}. In the presence of FM $J_1$ and/or $J_1'$, an intermediate phase (chiral II phase) appears between the high-field and low-field coplanar states. Furthermore, when the structure-change boundary becomes close to the upper critical field of the hedgehog-lattice phase, additional phases, the lower-temperature chiral I and higher-temperature chiral I' phases, show up in the vicinity of the upper boundary of the hedgehog-lattice phase.  
Here, we emphasize again that the chiral I and II phases as well as the in-field hedgehog-lattice phase exhibit nonzero total chirality $\mbox{\boldmath $\chi$}^{\rm T}$. Since $\mbox{\boldmath $\chi$}^{\rm T}$ and the magnetization $m$ show different field dependences (see the fourth and fifth panels from the top in Figs. \ref{fig:Hdep} and \ref{fig:HTdep}), the chirality-driven anomalous Hall signal could be distinguished from the usual anomalous Hall signal directly connected to the magnetization.

\section{Spin structures in in-field phases}
In this section, we will discuss the magnetic structures of the in-field phases all of which are quadruple-${\bf Q}$ $(\frac{1}{2}, \frac{1}{2}, \frac{1}{2})$ states with respect to the $S^xS^y$ spin component. It will be shown that the $S^z$ components in the hedgehog-lattice phase also exhibit a quadruple-${\bf Q}$ $(\frac{1}{2}, \frac{1}{2}, \frac{1}{2})$ structure, whereas those in the chiral I and chiral I' phases exhibit double-${\bf Q}$ $(\frac{1}{2}, \frac{1}{2}, \frac{1}{2})$ structures. 
Such a quadruple-${\bf Q}$ magnetic structure in the hedgehog-lattice phase turns out to be tetragonal-symmetric. As we will explain below, in the chiral I and chiral I' phases, a similar tetragonal-symmetric structure emerges, but the rotational symmetry in the plane perpendicular to the tetragonal axis is broken due to the double-${\bf Q}$ $S^z$ ordering, resulting in an orthorhombic-symmetric spin state. In the canted collinear, canted coplanar, and chiral II phases, the spin states are cubic-symmetric, tetragonal-symmetric, and orthorhombic-symmetric, respectively.

\begin{figure}[t]
\begin{center}
\includegraphics[width=\columnwidth]{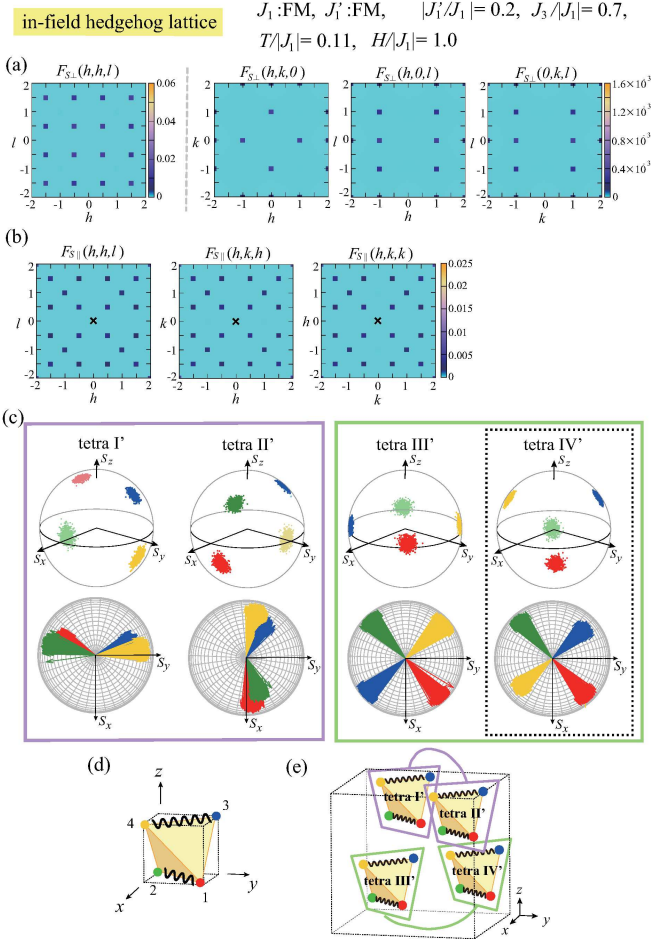}
\caption{Spin structure in the in-field hedgehog-lattice phase obtained at $T/|J_1|=0.11$ and $H/|J_1|=1.0$ for $|J_1'/J_1|=0.2$ and $J_3/|J_1|=0.7$ with FM $J_1$ and $J_1'$. (a) $F_{S\perp}({\bf q})$ in the $(h,h,l)$, $(h,k,0)$, $(h,0,l)$, and $(0,k,l)$ planes (from left to right) obtained for $L=12$ and (b) associated $F_{S\parallel}({\bf q})$ in the $(h,h,l)$, $(h,k,h)$, and $(h,k,k)$ planes (from left to right). In $F_{S\parallel}({\bf q})$, the high-intensity trivial peak at ${\bf q}=0$ indicated by a cross, which corresponds to $m^2$, has been removed. (c) MC spin snapshots mapped onto a unit sphere (upper panels) and their projection onto the $S^xS^y$ plane (lower panels), where the notations of the color and tetra's are the same as those in Fig. \ref{fig:hedgehog_snapmap} (a) and a monopole tetrahedron is enclosed by black dots. (d) Tetragonal-symmetric pairing pattern in each tetrahedron, where wavy lines denote pair bonds. (e) Tetragonal-symmetric distribution of the two different types of tetrahedra [tetra's enclosed by purple and green boxes in (c)] within the cubic unit cell (for details, see the text). 
\label{fig:infieldHL}}
\end{center}
\end{figure}

In the subsequent subsections, Figs. \ref{fig:infieldHL}-\ref{fig:cantcollinear} are obtained for $|J_1'/J_1|=0.2$ with FM $J_1$ and $J_1'$, whereas Figs. \ref{fig:cantcoplanar} and \ref{fig:chiral2} are obtained for $|J_1'/J_1|=0.6$ with AFM $J_1$ and FM $J_1'$. Since in all the ordered states, the spin structure factors for the $S^xS^y$ component $F_{S\perp}({\bf q})$ in the $(h,h,l)$, $(h,k,h)$, and $(h,k,k)$ planes exhibit almost the same Bragg-peak patterns, only the $(h,h,l)$ plane is shown in each (a) of Figs. \ref{fig:infieldHL}-\ref{fig:chiral2}. MC spin snapshots shown in Figs. \ref{fig:infieldHL}-\ref{fig:cantcollinear} are obtained by taking a short-time average over 10 MC steps to reduce the thermal noise. 
%
\subsection{In-field hedgehog-lattice phase}
Figure \ref{fig:infieldHL} shows the spin structure of the in-field hedgehog lattice in the FM $J_1$ and $J_1'$ case where the monopoles are formed on the large tetrahedra. Although the result below is essentially the same as that in the AFM $J_1$ and $J_1'$ case where the monopoles are formed on the small tetrahedra \cite{hedgehog_AK_prb_21}, here, we will discuss it for completeness. As one can see from Figs. \ref{fig:infieldHL} (a) and (b), the in-field hedgehog lattice is a quadruple-${\bf Q}$ $(\frac{1}{2}, \frac{1}{2}, \frac{1}{2})$ state with respect to both the $S^xS^y$ and $S^z$ spin components and is also characterized by additional $(1,0,0)$ and $(0,1,1)$-type Bragg peaks in $F_{S\perp}({\bf q})$. Although in Fig. \ref{fig:infieldHL} (b), $(1,1,1)$-type Bragg reflections can also be found, they are trivial ones stemming mainly from the uniform magnetization. In Fig. \ref{fig:infieldHL} (a), the $(1,0,0)$-type Bragg peaks appear at $(1,0,0)$ and $(0,1,0)$, while not at $(0,0,1)$, so that the spin state is tetragonal-symmetric in the sense that among the cubic-symmetric families of $(1,0,0)$, $(0,1,0)$, and $(0,0,1)$, only one direction ($z$-direction in the case of Fig. \ref{fig:infieldHL}) is special. Such a situation is also the case for the $(0,1,1)$-type Bragg reflections. Compared with the main peaks at $(\pm\frac{1}{2}, \pm\frac{1}{2}, \pm\frac{1}{2})$, these $(1,0,0)$ and $(0,1,1)$-type Bragg peaks show very weak intensity [see the color-bar ranges in Fig. \ref{fig:infieldHL} (a)].

We will next discuss how the tetragonal symmetry looks like in the real-space spin structure. Figure \ref{fig:infieldHL} (c) shows the associated spin snapshots mapped onto a unit sphere. One can see that on each large tetrahedron, spins belonging to the sublattices 1 and 2 (3 and 4) which, respectively, correspond to red and green  (blue and yellow) symbols in Fig. \ref{fig:infieldHL} (c) are paired up and their $S^xS^y$ components are almost collinear. As illustrated in Fig. \ref{fig:infieldHL} (d), bonds connecting the paired sublattices are stacking along $z$-axis, so that the spin configuration on each tetrahedron is tetragonal-symmetric with respect to $z$-axis. In the lower panels of Fig. \ref{fig:infieldHL} (c), one notices that on tetra's I' and II' (III' and IV') enclosed by a purple (green) box, the $S^xS^y$ components of paired spins are ferromagnetically (antiferromagnetically) aligned. As shown in Fig. \ref{fig:infieldHL} (e), these two types of tetrahedra are stacking along $z$-axis within the cubic unit cell. Thus, the spin state is tetragonal-symmetric at the level of the cubic unit cell as well as each tetrahedron. Although we have focused on the large tetrahedra, the small tetrahedra also possess the same tetragonal-symmetric real-space structure.   
As will be discussed below, the tetragonal symmetry of this kind can also be seen in the chiral I, chiral I', and canted coplanar phases, although in the chiral I and chiral I' phases, the spin-state symmetry is further reduced in the plane perpendicular to the tetragonal axis.

\subsection{Chiral I and chiral I' phases}
\begin{figure}[t]
\begin{center}
\includegraphics[width=\columnwidth]{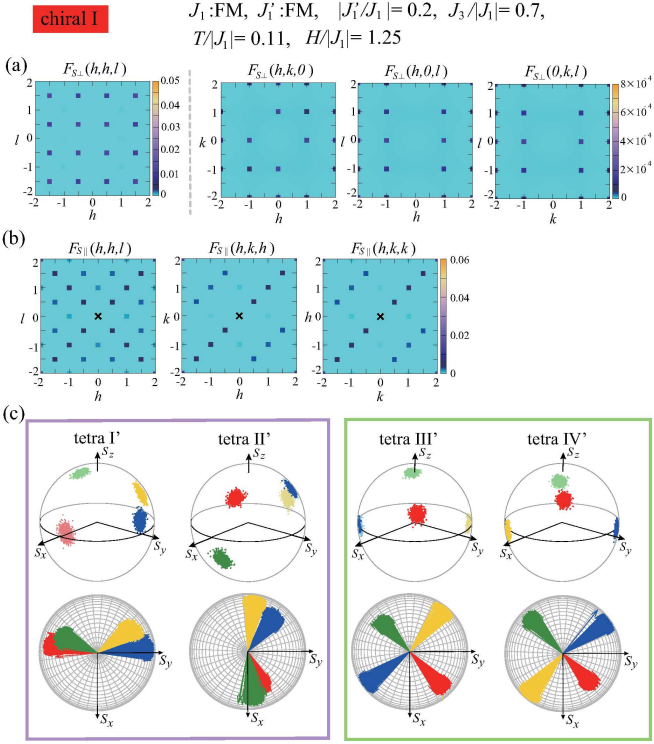}
\caption{Spin structure in the chiral I phase obtained at $T/|J_1|=0.11$ and $H/|J_1|=1.25$ for $|J_1'/J_1|=0.2$ and $J_3/|J_1|=0.7$ with FM $J_1$ and $J_1'$. (a) $F_{S\perp}({\bf q})$ in the $(h,h,l)$, $(h,k,0)$, $(h,0,l)$, and $(0,k,l)$ planes (from left to right) obtained for $L=12$ and (b) associated $F_{S\parallel}({\bf q})$ in the $(h,h,l)$, $(h,k,h)$, and $(h,k,k)$ planes (from left to right), where in (b), the high-intensity trivial peak at ${\bf q}=0$ indicated by a cross has been removed. (c) MC snapshots of spins mapped onto a unit sphere (upper panels) and their projection onto the $S^xS^y$ plane (lower panels), where the notations are the same as those in Fig. \ref{fig:infieldHL} (c). 
\label{fig:chiral1L}}
\end{center}
\end{figure}

In this subsection, we will first discuss the chiral I phase, and then, explain the difference between the chiral I and chiral I' phases whose spin structures look quite similar.
Figure \ref{fig:chiral1L} shows the spin structure in the chiral I phase. In $F_{S\parallel}({\bf q})$ shown in Fig. \ref{fig:chiral1L} (b), one can see Bragg peaks at $\pm{\bf Q}_1=\pm(\frac{1}{2}, \frac{1}{2}, \frac{1}{2})$ and $\pm{\bf Q}_4=\pm(\frac{1}{2}, \frac{1}{2}, \frac{-1}{2})$ but not at the remaining two ${\bf Q}_2$ and ${\bf Q}_3$, so that the $S^z$ component forms a double-${\bf Q}$ $(\frac{1}{2}, \frac{1}{2}, \frac{1}{2})$ structure, whereas the $S^xS^y$ component forms a quadruple-${\bf Q}$ structure [see the left panel of Fig. \ref{fig:chiral1L} (a)]. In $F_{S\perp}({\bf q})$, in addition to the quadruple-${\bf Q}$ $(\frac{1}{2}, \frac{1}{2}, \frac{1}{2})$ Bragg peaks, weak-intensity $(1,0,0)$- and $(0,1,1)$-type peaks appear, as in the case of the in-field hedgehog lattice [note that the $(1,1,0)$ peak is hardly visible in the $(h,h,l)$ plane shown in the left panel of Fig. \ref{fig:chiral1L} (a), because its intensity is considerably weak as indicated by the right color-bar range in Fig. \ref{fig:chiral1L} (a)]. Since these $(1,0,0)$- and $(0,1,1)$-type Bragg peaks are absent only at one of the cubic-symmetric $(1,0,0)$ and $(0,1,1)$ families [see the right three panels in Fig. \ref{fig:chiral1L}], at first sight, the spin state looks tetragonal-symmetric with $z$-axis being characteristic. 
Compared with $F_{S\perp}({\bf q})$ in the in-field hedgehog-lattice phase (see Fig. \ref{fig:infieldHL}), however, $F_{S\perp}({\bf q})$ in the chiral I phase additionally shows Bragg peaks at $\pm(1,1,0)$ but not at $\pm(1,-1,0)$. Such a nonequivalence between $(1,1,0)$ and $(1,-1,0)$ has been confirmed in $F_{S\parallel}({\bf q})$ as well. 
This ordering vector $(1,1,0)$ distinguished from $(1,-1,0)$ is associated with the double-${\bf Q}$ ordering vectors ${\bf Q}_1$ and ${\bf Q}_4$ via the relation ${\bf Q}_1+{\bf Q}_4=(1,1,0)$. We have confirmed this association for different MC snapshots. 
Since $(1,1,0)$ and $(1,-1,0)$ are not equivalent any more, the rotational symmetry in the plane perpendicular to the symmetry axis ($z$-axis) is broken, which is reflected in nonvanishing $P_{\perp 2}$ (see the third panels from the top in Figs. \ref{fig:Hdep} and \ref{fig:HTdep}). As these orthogonal two directions are not equivalent, the spin state turns out to be orthorhombic-symmetric.

%
%
\begin{figure}[t]
\begin{center}
\includegraphics[width=\columnwidth]{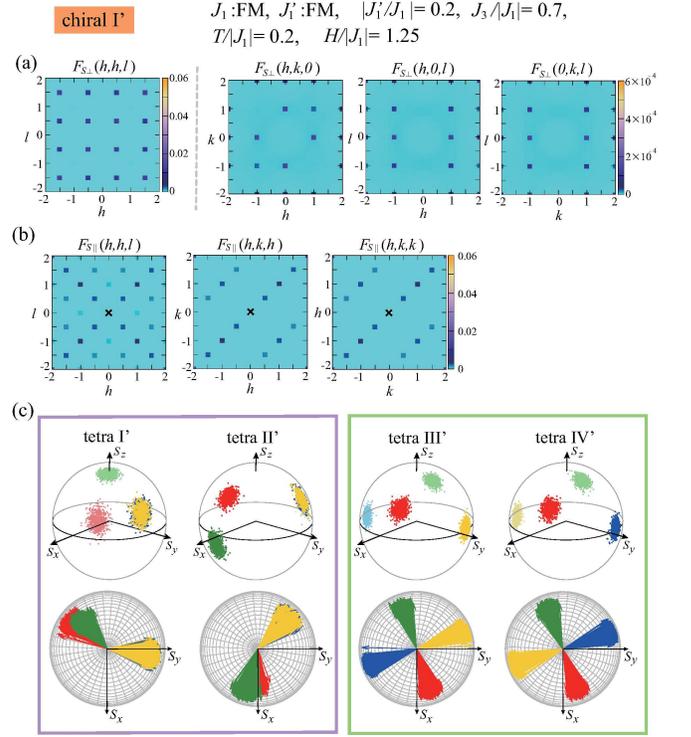}
\caption{Spin structure in the chiral I' phase obtained at $T/|J_1|=0.2$ and $H/|J_1|=1.25$ for $|J_1'/J_1|=0.2$ and $J_3/|J_1|=0.7$ with FM $J_1$ and $J_1'$. (a) $F_{S\perp}({\bf q})$ in the $(h,h,l)$, $(h,k,0)$, $(h,0,l)$, and $(0,k,l)$ planes (from left to right) obtained for $L=12$ and (b) associated $F_{S\parallel}({\bf q})$ in the $(h,h,l)$, $(h,k,h)$, and $(h,k,k)$ planes (from left to right). (c) MC spin snapshots mapped onto a unit sphere (upper panels) and their projection onto the $S^xS^y$ plane (lower panels), where the notations are the same as those in Figs. \ref{fig:infieldHL} (c) and \ref{fig:chiral1L} (c). 
\label{fig:chiral1H}}
\end{center}
\end{figure}
%
As one can see from MC snapshots shown in Fig. \ref{fig:chiral1L} (c), the real-space spin structure is quite similar to that of the in-field hedgehog lattice. On each tetrahedron, the four sublattices are paired up into two such that spins belonging to each pair tend to orient along almost the same axis within the $S^xS^y$ plane, and on tetra's I' and II' (III' and IV'), the $S^xS^y$ components of paired-sublattice spins are ferromagnetically (antiferromagnetically) aligned. Furthermore, in the three-dimensional spin space, four spins on each tetrahedron are pointing in different directions. Thus, the real-space structure possesses almost the same kind of tetragonal symmetry as that in Figs. \ref{fig:infieldHL} (d) and (e). Although the further symmetry reduction discussed above can hardly be seen in the real-space spin structure, it is reflected in the chirality sector more clearly. 

In the chirality sector [see Fig. \ref{fig:Cq2comp} (b) in Appendix A], the structure factor $F_C({\bf q})$ in the chiral I phase exhibits Bragg peaks at $\pm{\bf Q}_1$,  $\pm{\bf Q}_2$, $\pm{\bf Q}_3$, and $\pm{\bf Q}_4$, but in contrast to the hedgehog-lattice phase where their intensities are basically the same [see Fig. \ref{fig:Cq2comp} (a) in Appendix A], the ${\bf Q}_1$ and ${\bf Q}_4$ intensities are relatively weaker than the ${\bf Q}_2$ and ${\bf Q}_3$ ones, reflecting the double-${\bf Q}$ nature of the $S^z$ spin component. As will be discussed in Sec VI, the direction of the total chirality vector $\mbox{\boldmath $\chi$}^{\rm T}$ is determined by the double-${\bf Q}$ ordering vectors, i.e., the rotational symmetry breaking in the plane perpendicular to the main symmetry axis.

Now that the spin structure in the chiral I phase is understood, we will next discuss the chiral I' phase which is clearly distinguished from the chiral I phase by the absence of the total chirality $\mbox{\boldmath $\chi$}^{\rm T}$ (compare the red and orange regions in the forth panels from the top in Figs. \ref{fig:Hdep} and \ref{fig:HTdep}). By comparing Figs. \ref{fig:chiral1H} (a) and (b) with Figs. \ref{fig:chiral1L} (a) and (b), one notices that the spin structure factors in the chiral I and chiral I' phases are qualitatively the same including the double-${\bf Q}$ nature for the $S^z$ component, suggestive of the spin-state symmetry of the same kind. Actually, the real-space structures in the chiral I and chiral I' phases shown in Figs. \ref{fig:chiral1L} (c) and \ref{fig:chiral1H} (c) look qualitatively the same. There is, however, a minor but significant difference in the spin configurations on the tetrahedra having sublattice pairs with ferromagnetically-aligned $S^xS^y$ components. In the upper panels of Fig. \ref{fig:chiral1H} (c), spins belonging to sublattices 3 and 4 [blue and yellow symbols in Fig. \ref{fig:chiral1H} (c)] are oriented in the same direction on tetra's I' and II'. This is in sharp contrast to the chiral I and in-field hedgehog-lattice phases where the four sublattice spins are pointing in different directions on each tetrahedron. 

Although the difference between the chiral I and chiral I' phases are hardly visible in the spin structure factors shown in Figs. \ref{fig:chiral1L} (a) and (b) and Figs. \ref{fig:chiral1H} (a) and (b), it can clearly be seen in the chirality structure factor [see the left panels of Figs. \ref{fig:Cq2comp} (b) and (c)]: in the chiral I' phase, the $(\frac{1}{2}, \frac{1}{2}, \frac{1}{2})$ and $(\frac{1}{2}, \frac{1}{2}, \frac{-1}{2})$ peaks disappear in $F_C({\bf q})$, and the $(\frac{1}{2}, \frac{1}{2}, \frac{1}{2})$ double-${\bf Q}$ nature is more remarkable.
%
\begin{figure}[t]
\begin{center}
\includegraphics[scale=0.38]{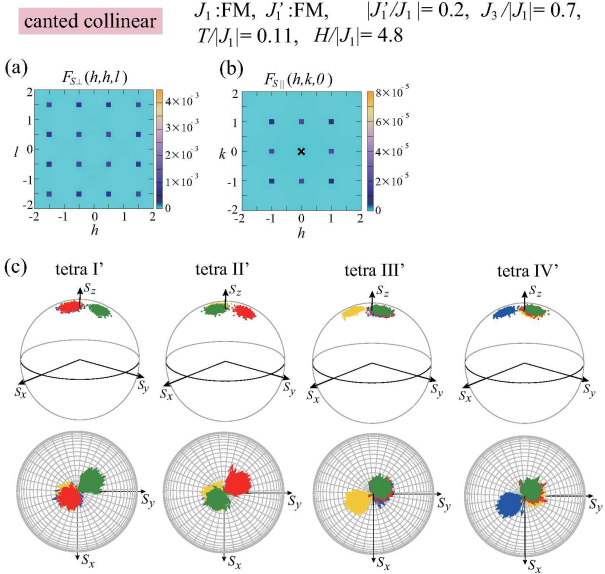}
\caption{Spin structure in the canted collinear phase obtained at $T/|J_1|=0.11$ and $H/|J_1|=4.8$ for $|J_1'/J_1|=0.2$ and $J_3/|J_1|=0.7$ with FM $J_1$ and $J_1'$. (a) $F_{S\perp}({\bf q})$ in the $(h,h,l)$ plane obtained for $L=12$ and (b) associated $F_{S\parallel}({\bf q})$ in the $(h,k,0)$ plane. (c) MC snapshots of spins mapped onto a unit sphere (upper panels) and their projection onto the $S^xS^y$ plane (lower panels), where the notations are the same as those in Figs. \ref{fig:infieldHL} (c), \ref{fig:chiral1L} (c), and \ref{fig:chiral1H} (c). \label{fig:cantcollinear}}
\end{center}
\end{figure}
%
\begin{figure}[t]
\begin{center}
\includegraphics[width=\columnwidth]{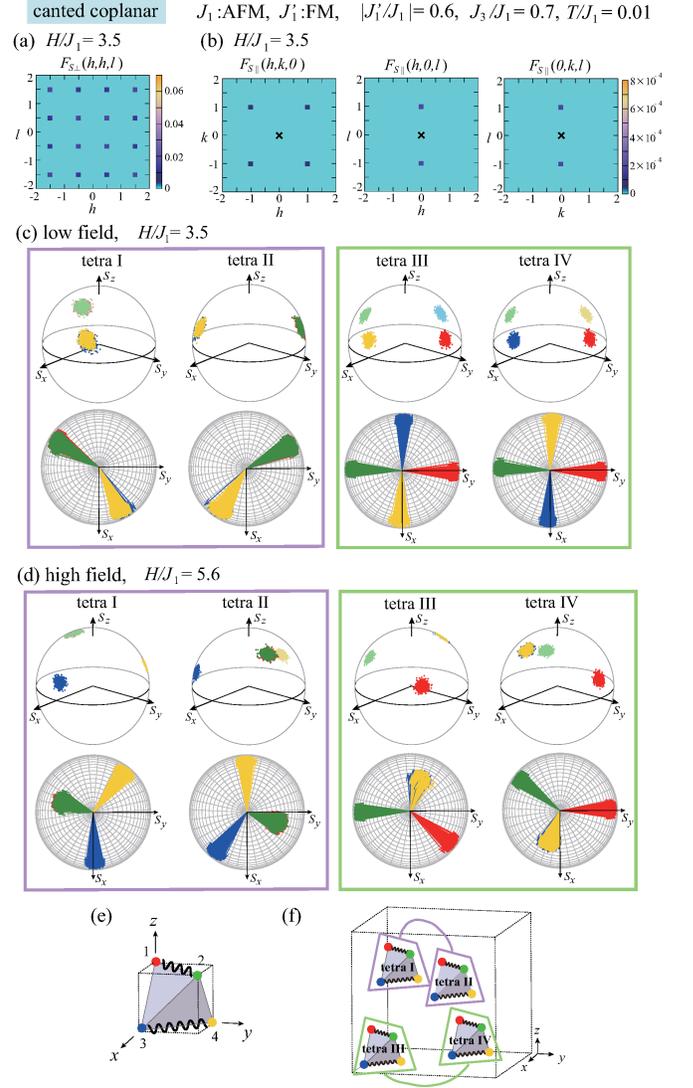}
\caption{Spin structure in the canted coplanar phase obtained at $T/J_1=0.01$ for $|J_1'/J_1|=0.6$ and $J_3/J_1=0.7$ with AFM $J_1$ and FM $J_1'$. (a) $F_{S\perp}({\bf q})$ in the $(h,h,l)$ plane obtained at $H/J_1=3.5$ for $L=12$ and (b) associated $F_{S\parallel}({\bf q})$ in the $(h,k,0)$, $(h,0,l)$, and $(0,k,l)$ planes (from left to right). (c) and (d) MC snapshots of spins mapped onto a unit sphere (upper panels) and their projection onto the $S^xS^y$ plane (lower panels) obtained at $H/J_1=3.5$ and 5.6, respectively, where the notations of the color and tetra's are the same as those in Fig. \ref{fig:hedgehog_snapmap} (b). (e) Tetragonal-symmetric pairing pattern in each tetrahedron, where wavy lines denote pair bonds. (f) Tetragonal-symmetric distribution of the two different types of tetrahedra [purple and green tetra's in (c) and (d)] within the cubic unit cell (for details, see the text).  
\label{fig:cantcoplanar}}
\end{center}
\end{figure}
%
\begin{figure}[t]
\begin{center}
\includegraphics[width=\columnwidth]{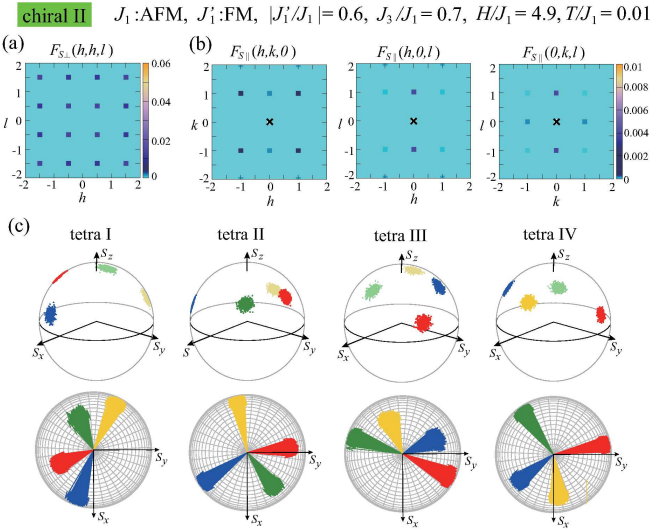}
\caption{Spin structure in the chiral II phase obtained at $T/J_1=0.01$ and $H/J_1=4.9$ for $|J_1'/J_1|=0.6$ and $J_3/J_1=0.7$ with AFM $J_1$ and FM $J_1'$. (a) $F_{S\perp}({\bf q})$ in the $(h,h,l)$ plane obtained for $L=12$ and (b) associated $F_{S\parallel}({\bf q})$ in the $(h,k,0)$, $(h,0,l)$, and $(0,k,l)$ planes (from left to right), where in (b), the high-intensity trivial peak at ${\bf q}=0$ indicated by a cross, which corresponds to $m^2$, has been removed. (c) MC snapshots of spins mapped onto a unit sphere (upper panels) and their projection onto the $S^xS^y$ spin plane (lower panels), where the notations are the same as those in Figs. \ref{fig:cantcoplanar} (c) and (d). 
\label{fig:chiral2}}
\end{center}
\end{figure}
%
\subsection{Canted collinear phase}
In the canted collinear phase, quadruple-${\bf Q}$ $(\frac{1}{2}, \frac{1}{2}, \frac{1}{2})$ Bragg peaks appear only in $F_{S\perp}({\bf q})$, and the structure of the $S^z$ component is characterized by the cubic-symmetric $(1,0,0)$- and $(0,1,1)$-type ordering vectors [see Figs. \ref{fig:cantcollinear} (a) and (b)]. Thus, the spin state is cubic-symmetric. Actually, as one can see from Fig. \ref{fig:cantcollinear} (c), all the four-sublattice spins are collinearly aligned in the $S^xS^y$ plane, and the sublattice pairing characteristic of the tetragonal-symmetric spin configuration does not occur. Four spins on each of large tetrahedra shown in Fig. \ref{fig:cantcollinear} (c) take a 3:1 configuration within the $S^xS^y$-plane collinear manifold, whereas those on each small tetrahedron take a 2:2 or 4:0 configuration (not shown here), reflecting the zero-field collinear spin structure [see Fig. 8 (a) in Ref. \cite{hedgehog_AK_prb_21}]. This phase is characterized by the $S^xS^y$-plane nematicity $P_{\perp 2}$ [see the third panel from the top in Fig. \ref{fig:Hdep} (a)]. 

\subsection{Canted coplanar phase}
Figure \ref{fig:cantcoplanar} shows the spin structure of the canted coplanar phase, where $(\frac{1}{2}, \frac{1}{2}, \frac{1}{2})$ Bragg peaks appear only in $F_{S\perp}({\bf q})$ like in the case of the canted collinear phase. As one can see from Fig. \ref{fig:cantcoplanar} (b), the canted coplanar state is tetragonal-symmetric because among cubic-symmetric families of $(1,0,0)$ and $(0,1,1)$, only $(0,0,1)$ and $(1,\pm 1,0)$ are picked up in $F_{S\parallel}({\bf q})$ with $z$-axis being characteristic. We note that in the canted coplanar phase, the tetragonal symmetry is reflected in the $S^z$ component, whereas in the in-field hedgehog-lattice phase, it is reflected in the $S^xS^y$ component. 
Within the canted coplanar phase, the spin structure factor looks qualitatively the same irrespective of the field strength, and thereby, the tetragonal symmetry remains unchanged. Nevertheless, the real-space spin structure in the high-field regime is slightly different from that in the low-field regime. 

Figures \ref{fig:cantcoplanar} (c) and (d) show the MC spin snapshots obtained in the low-field and high-field regions of the canted coplanar phase, respectively. The common feature of the two is that the four sublattice 1, 2, 3, and 4 are paired up into two on each tetrahedron as indicated in Fig. \ref{fig:cantcoplanar} (e) and that the four tetrahedra I, II, III, and IV are classified into two each of which is enclosed by a purple or green box in Figs. \ref{fig:cantcoplanar} (c) and (d). As shown in Fig. \ref{fig:cantcoplanar} (f), both the paired sublattices and the two classes of tetrahedra are stacking along $z$-axis. The difference between the low-field and high-field structures is that in the high-field region, spins belonging to one paired sublattices on each tetrahedron are pointing in the same direction, while not in the low-field region. Nevertheless, we cannot find a qualitative difference between the two in the physical quantities, so that we call these two structures with the single term of "canted coplanar".

\subsection{Chiral II phase}
The spin structure in the chiral II phase possesses the features of the neighboring low-field and high-field coplanar phases. As shown in Figs. \ref{fig:chiral2} (a) and (b), the Bragg peak patterns are basically the same as those in the canted coplanar phase shown in Figs. \ref{fig:cantcoplanar} (a) and (b) except the additional weak peaks at $\pm(0,1,0)$ in $F_{S\parallel}({\bf q})$. Since the intensities of the main $(0,0,1)$ and additional $(0,1,0)$ peaks are clearly different and the $(1,0,0)$ component is absent, the spin state could be categorized as orthorhombic. Actually, in the real-space spin structure shown in Fig. \ref{fig:chiral2} (c), one cannot see clear sublattice pairings characteristic of the tetragonal-symmetric spin structures. On the other hand, Fig. \ref{fig:chiral2} (c) could be viewed as an intermediate state between Figs. \ref{fig:cantcoplanar} (c) and (d): starting from the high-field coplanar structure in Fig. \ref{fig:cantcoplanar} (d), the paired sublattices with the same spin orientation are split into two to reconstruct the spin configuration such that the low-field structure exemplified by Fig. \ref{fig:cantcoplanar} (c) is obtained. In such a split state, i.e., the chiral II phase, the spin directions for the four sublattices are different on each tetrahedron, as in the in-field hedgehog-lattice and chiral I phases. Interestingly, the total chirality $\mbox{\boldmath $\chi$}^{\rm T}$ is nonzero only in these phases where the four spins on each of all the tetrahedra are pointing in the different directions in the three-dimensional spin space.   

In the chiral II phase, the local chirality summed over the four triangles on a tetrahedron $\chi({\bf R}_l) = \sum_{i,j,k \in l \, {\rm th} \, {\rm tetra}} \chi_{ijk}$ vanishes, so that Bragg peaks are trivially absent in the associated chirality structure factor $F_C({\bf q})$. 
Nevertheless, with the nonzero $\mbox{\boldmath $\chi$}^{\rm T}$, we call this state ''chiral''.

\section{Field-induced total chirality}
Now that the spin structures in all the in-field phases are clarified, here, we will discuss the origin of the field-induced total chirality $\mbox{\boldmath $\chi$}^{\rm T}$ associated with the chirality-driven anomalous Hall effect. 
Since as one can see from Eq. (\ref{eq:total_chi}), the chirality ''vector'' $\mbox{\boldmath $\chi$}^{\rm T}$ is obtained by summing up the contributions from all the tetrahedra, we shall start from the fundamental properties of the chirality vector for a single tetrahedron $\mbox{\boldmath $\chi$}^{\rm T,tetra}=\langle \sum_{ i,j,k \in {\rm tetra}} \chi_{ijk} \, \hat{n}_{ijk} \, \rangle$ with $\chi_{ijk}={\bf S}_i \cdot ({\bf S}_j \times {\bf S}_k)$.

\subsection{Chirality for a tetrahedron}
\begin{figure}[t]
\begin{center}
\includegraphics[width=\columnwidth]{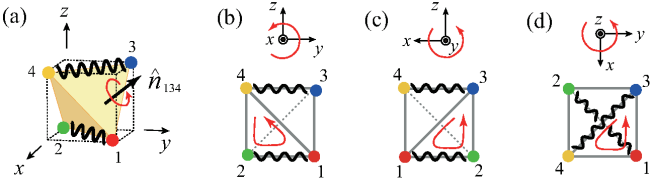}
\caption{Chirality for a tetrahedron. (a) A tetrahedron having a tetragonal-symmetric sublattice pairing, and (b)-(d) its projections onto the two-dimensional planes perpendicular to $x$-, $y$-, and $z$-axes, respectively, where a red circular arrow denotes the direction in which the spin chirality $\chi_{ijk}$ is deﬁned and other notations are the same as those in Fig. \ref{fig:infieldHL} (d).  
\label{fig:tetraconf}}
\end{center}
\end{figure}
%
For the large tetrahedron shown in Fig. \ref{fig:tetraconf} (a), the surface normals $\hat{n}_{ijk}$ are given by $\hat{n}_{134}=(1,1,1)/\sqrt{3}$,  $\hat{n}_{142}=(1,-1,-1)/\sqrt{3}$, $\hat{n}_{243}=(-1,-1,1)/\sqrt{3}$, and $\hat{n}_{123}=(-1,1,-1)/\sqrt{3}$, so that the chirality vector for the large tetrahedron is given by
\begin{equation}
\mbox{\boldmath $\chi$}^{\rm T,tetra} = \frac{1}{\sqrt{3}}\left( \begin{array}{c}
\chi_{134} + \chi_{142} - \chi_{243} - \chi_{123} \\
\chi_{134} - \chi_{142} - \chi_{243} + \chi_{123} \\
\chi_{134} - \chi_{142} + \chi_{243} - \chi_{123} 
\end{array} \right) .
\end{equation}
Noting that $\chi_{ijk}= - \chi_{ikj}$, one finds that the $x$-component of $\mbox{\boldmath $\chi$}^{\rm T,tetra}$ is none other than the total chirality summed over the four $yz$-plane triangles yielded when the tetrahedron is projected onto the $yz$-plane [see Fig. \ref{fig:tetraconf} (b)]. Such a situation is also the case for the $xz$- and $xy$-projections [see Figs. \ref{fig:tetraconf} (c) and (d)], suggesting that the spin configuration on the projected layer of one-tetrahedron width is essential for the chirality vector corresponding to the emergent fictitious field. Below in this subsection, we will discuss how the spin configuration on a tetrahedron is reflected in the chirality in the projected two-dimensional planes.

In the canted collinear phase shown in Fig. \ref{fig:cantcollinear} (c), all the spins are in the same spin plane, so that the local scalar chirality $\chi_{ijk}$ trivially vanishes and thereby, $\mbox{\boldmath $\chi$}^{\rm T,tetra} =0$. Thus, the question is how the tetragonal-symmetric spin configuration on a tetrahedron, which can commonly be seen in most of the in-field phases, is reflected in the chirality in the projected two-dimensional planes. 
As one can see from Figs. \ref{fig:tetraconf} (a)-(d), the sublattice pairs stacking along the tetragonal-symmetric $z$-axis [see the wavy lines in Fig. \ref{fig:tetraconf} (a)] are arranged in parallel in the projected planes perpendicular to $x$ and $y$ axes [see Figs. \ref{fig:tetraconf} (b) and (c)], whereas arranged diagonally in the plane perpendicular to the tetragonal symmetry axis, i.e., $z$-axis [see Fig. \ref{fig:tetraconf} (d)]. 
Such a difference in the arrangement pattern turns out to be important for whether $\mbox{\boldmath $\chi$}^{\rm T,tetra}$ eventually vanishes or not.
%
\begin{figure}[t]
\begin{center}
\includegraphics[width=\columnwidth]{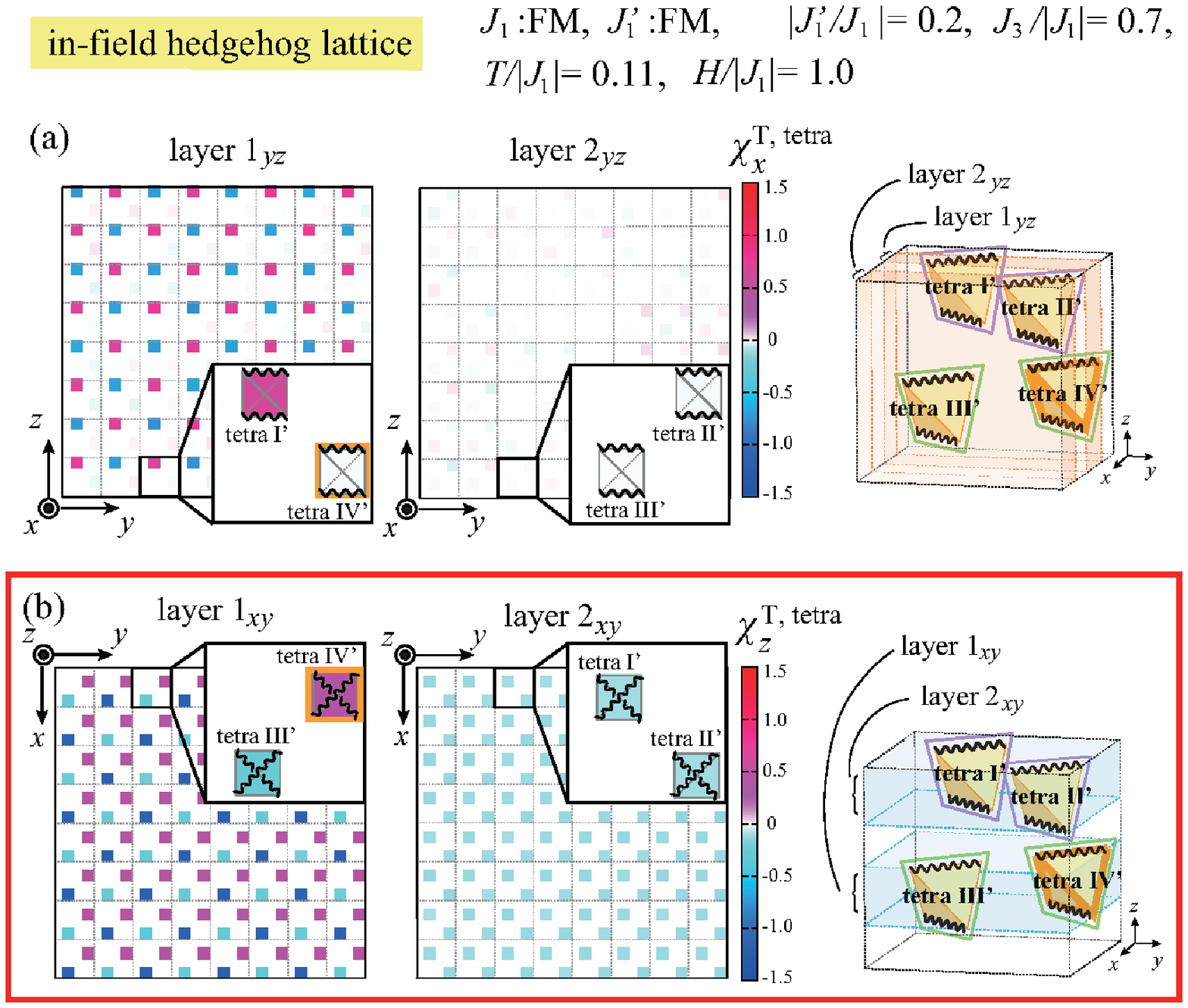}
\caption{Chirality distributions in the projected two-dimensional planes in the in-field hedgehog-lattice phase obtained at $T/|J_1|=0.11$ and $H/|J_1|=1.0$ for $|J_1'/J_1|=0.2$ and $J_3/|J_1|=0.7$ with FM $J_1$ and $J_1'$. (a) and (b) Layer-resolved distributions of the chirality calculated from the MC snapshot in Fig. \ref{fig:infieldHL}, where short-time average over 200 MC steps has been made to reduce the thermal noise. In (a) [(b)], $\chi^{\rm T,tetra}_x$'s ($\chi^{\rm T,tetra}_z$'s) on the $yz$- ($xy$-)plane layers deﬁned in a right panel are shown, where the inset shows a zoomed view of the projected cubic unit cell. In the zoomed view, a wavy line connects paired sublattices, and a tetrahedra outlined by a thick orange box (tetra IV') corresponds to the monopole tetrahedron. Nonzero contributions to the total chirality come from the layers enclosed by a red box, which in the present case, correspond to the $xy$-plane layers perpendicular to the tetragonal symmetry axis shown in (b).  
\label{fig:chidis_infieldHL}}
\end{center}
\end{figure}
%
\begin{figure}[t]
\begin{center}
\includegraphics[width=\columnwidth]{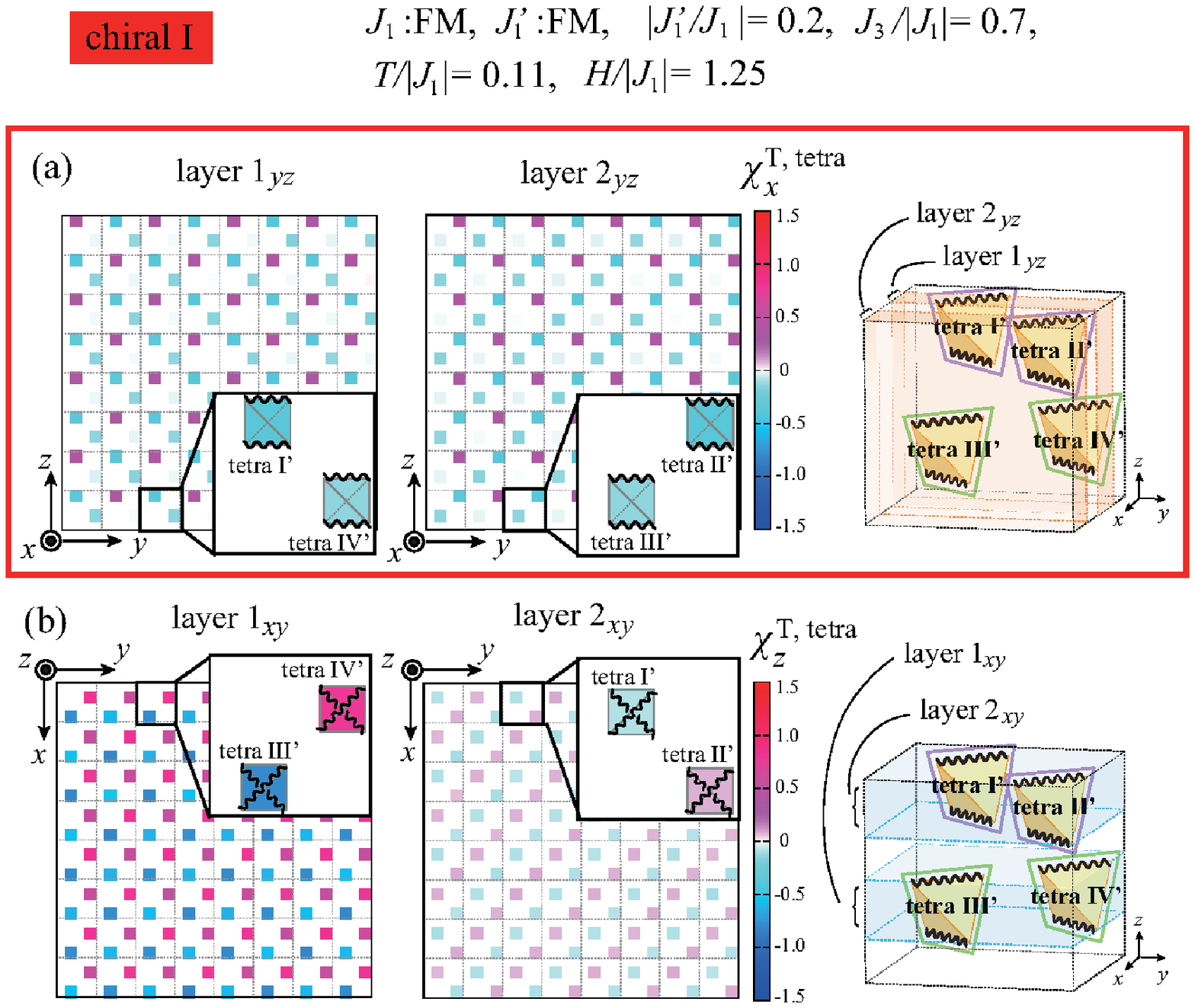}
\caption{Chirality distributions in the projected two-dimensional planes in the chiral I phase obtained at $T/|J_1|=0.11$ and $H/|J_1|=1.25$ for $|J_1'/J_1|=0.2$ and $J_3/|J_1|=0.7$ with FM $J_1$ and $J_1'$. (a) and (b) Layer-resolved distributions of the chirality calculated from the MC snapshot in Fig. \ref{fig:infieldHL}, where short-time average over 200 MC steps has been made to reduce the thermal noise. In (a) [(b)], $\chi^{\rm T,tetra}_x$'s ($\chi^{\rm T,tetra}_z$'s) on the $yz$- ($xy$-)plane layers defined in a right panel are shown, and other notations are the same as those in Fig. \ref{fig:chidis_infieldHL}.\label{fig:chidis_chiral1L}}
\end{center}
\end{figure}
%
\begin{figure}[t]
\begin{center}
\includegraphics[width=\columnwidth]{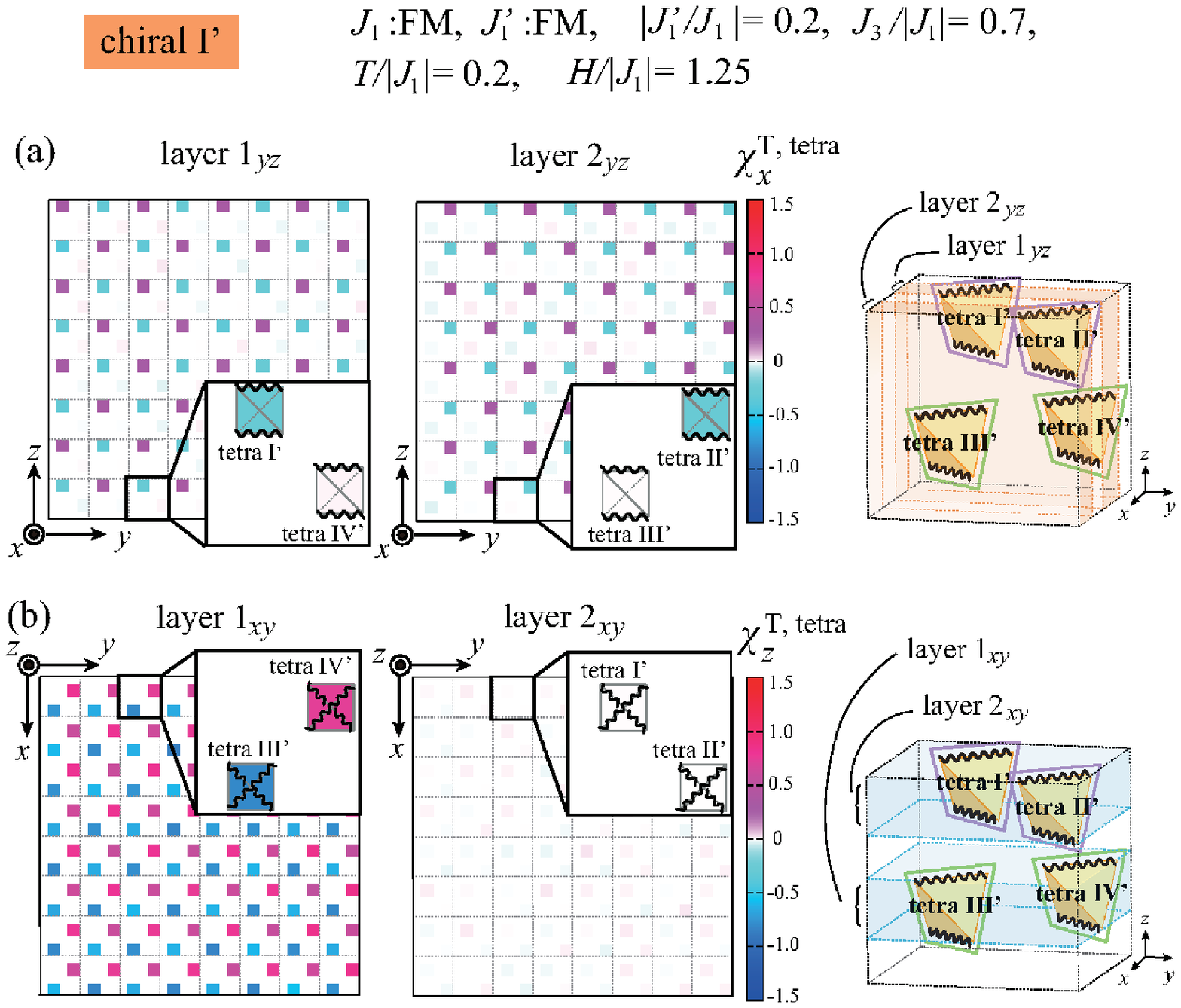}
\caption{Chirality distributions in the projected two-dimensional planes in the chiral I phase obtained at $T/|J_1|=0.2$ and $H/|J_1|=1.25$ for $|J_1'/J_1|=0.2$ and $J_3/|J_1|=0.7$ with FM $J_1$ and $J_1'$. (a) and (b) Layer-resolved distributions of the chirality calculated from the MC snapshot in Fig. \ref{fig:infieldHL}, where short-time average over 200 MC steps has been made to reduce the thermal noise. In (a) [(b)], $\chi^{\rm T,tetra}_x$'s ($\chi^{\rm T,tetra}_z$'s) on the $yz$- ($xy$-)plane layers deﬁned in a right panel are shown, and other notations are the same as those in Fig. \ref{fig:chidis_infieldHL}.  \label{fig:chidis_chiral1H}}
\end{center}
\end{figure}
%
\begin{figure}[t]
\begin{center}
\includegraphics[width=\columnwidth]{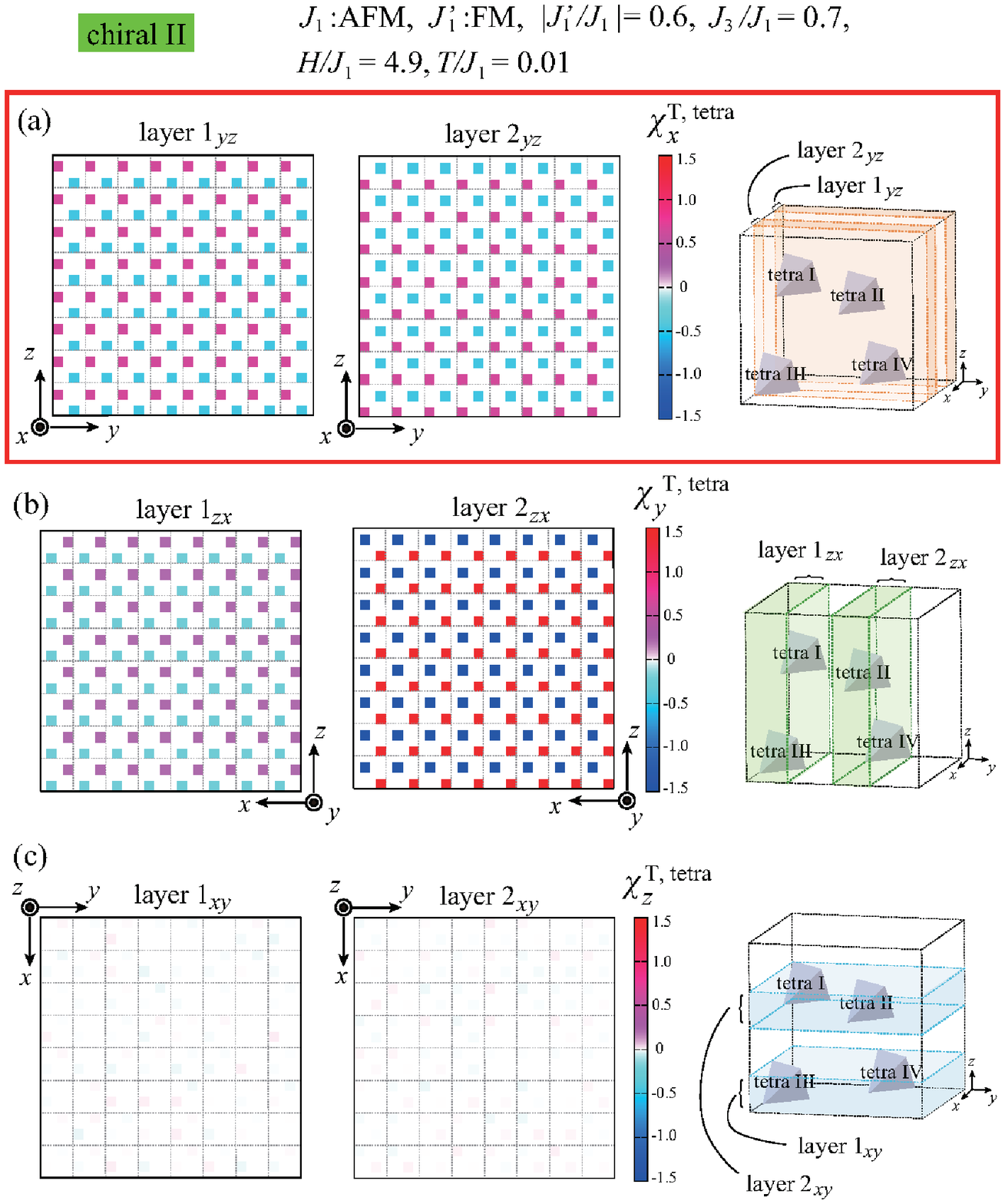}
\caption{Chirality distributions in the projected two-dimensional planes in the chiral II phase obtained at $T/J_1=0.01$ and $H/J_1=4.9$ for $|J_1'/J_1|=0.6$ and $J_3/J_1=0.7$ with AFM $J_1$ and FM $J_1'$. (a) and (b) Layer-resolved distributions of the chirality calculated from the MC snapshot in Fig. \ref{fig:infieldHL}, where short-time average over 200 MC steps has been made to reduce the thermal noise. In (a)-(c), $\chi^{\rm T,tetra}_x$'s, $\chi^{\rm T,tetra}_y$'s, and $\chi^{\rm T,tetra}_z$'s on the $yz$-, $zx$-, and $xy$-plane layers deﬁned in right panels are shown, respectively. Nonzero contributions to the total chirality come from the layers enclosed by a red box, which in the present case, correspond to the $yz$-plane layers shown in (a).\label{fig:chidis_chiral2}}
\end{center}
\end{figure}
%

In an applied magnetic field, a spin can be expressed with the use of the uniform magnetization $m$ as ${\bf S}_i = (m + \delta S^z_i)\hat{z} + \delta {\bf S}^\perp_i$. In the case of the in-field hedgehog lattice shown in Fig. \ref{fig:infieldHL} (c), $S^xS^y$ components of paired spins are almost collinear, so that for simplicity, here, we assume that $\delta {\bf S}^\perp_i$'s and $\delta S^z_i$'s for the paired sublattices are parallel and antiparallel (antiparallel and parallel), respectively, on tetra's I' and II' (III' and IV'). In the low-field coplanar case shown in Fig. \ref{fig:cantcoplanar} (c), such a situation is also the case for $\delta {\bf S}^\perp_i$, whereas $\delta S^z_i$'s are uniform. By using such constraints on $\delta {\bf S}_i$, we can perform reference calculations of the chirality on a projected tetrahedron for three types of arrangement patterns of the paired sublattices shown in Figs. \ref{fig:tetraconf} (b)-(d).

It turns out that in the in-field hedgehog-lattice phase, the tetrahedron chirality is zero or an odd function of $\delta S^\mu_i$ for the parallel arrangements in Figs. \ref{fig:tetraconf} (b) and (c), whereas it involves an even function of $\delta S^\mu_i$ for the diagonal arrangement in Fig. \ref{fig:tetraconf} (d) [see Eqs. (\ref{eq:tetra_chi_type1}) and (\ref{eq:tetra_chi_type2}) in Appendix B]. As will be demonstrated in the next subsection, when summed over the whole two-dimensional plane, the chirality is nonvanishing only for the diagonal pairing, i.e., in the tetragonal-symmetric direction, while it is zero from the beginning or completely canceled out for the parallel pairing, i.e., in the remaining two directions. In the low-field coplanar phase, due to the uniform $\delta S^z_i$, $\mbox{\boldmath $\chi$}^{\rm T,tetra}$ is zero for the parallel pairings and becomes an odd function of $\delta S^\mu_i$ for the diagonal pairing, whereas in the high-field coplanar phase, it is an odd function for the parallel pairings and zero for the diagonal pairing [see Eq. (\ref{eq:tetra_chi_type3}) in Appendix B]. In both cases, the nonzero local tetrahedron chirality is completely canceled out by a contribution from a counter tetrahedron, reflecting the fact that it is an odd function. In the chiral I, chiral I', and chiral II phases, whether the cancellation occurs or not is not so trivial. In the next subsection, we will discuss the distribution of the chirality over the whole projected two-dimensional plane in the chiral phases including the in-field hedgehog lattice. For completeness, the chirality distribution in the canted coplanar phase is also shown in Appendix C.

\subsection{Real-space chirality distributions in the chiral phases}
Figure \ref{fig:chidis_infieldHL} shows the real-space chirality distributions in the projected two-dimensional planes in the in-field hedgehog-lattice phase with the tetragonal-symmetric spin structure with respect to $z$-axis. Since the quadruple-${\bf Q}$ $(\frac{1}{2},\frac{1}{2},\frac{1}{2})$ state consists of the alternating array of two types of cubic unit cells, the layers shown in each of Figs. \ref{fig:chidis_infieldHL} (a) and (b) are stacking alternately. One can see from Fig. \ref{fig:chidis_infieldHL} (a) that on the layers perpendicular to $x$-axis (non-symmetric axis) where the paired sublattices are arranged in parallel, the local chiralities $\chi^{\rm T,tetra}_x$'s for most of the tetrahedra vanish due to the reason explained in the previous subsection. Although an exceptional tetrahedron, e.g., tetra I' in the zoomed view in Fig. \ref{fig:chidis_infieldHL} (a), has a nonzero $\chi^{\rm T, tetra}_x$, it is completely cancelled out by the contribution from the counter tetrahedron in the neighboring cubic unit cell, so that the net chirality vanishes on any $yz$-plane layer. Such a situation is also the case on the layers perpendicular to $y$-axis, and thus, the total chirality $\chi^{\rm T, tetra}_y$ also vanishes.

By contrast, on the layers perpendicular to $z$-axis (tetragonal symmetry axis) where the paired spins are arranged diagonally, not only the local chirality $\chi^{\rm T,tetra}_z$ but also the net chirality becomes nonzero. Thus, in the tetragonal-symmetric in-field hedgehog lattice phase, the total chirality $\mbox{\boldmath $\chi$}^{\rm T}$ is induced along the tetragonal symmetry axis. On the layer with the monopoles [layer $1_{xy}$ in Fig. \ref{fig:chidis_infieldHL} (b)], the monopole and antimonopole tetrahedra yield $\chi^{\rm T, tetra}_z$'s of the same sign, and their contributions are cancelled out by those of the opposite sign from other neighboring tetrahedra. Although the cancellation is incomplete resulting in a nonzero small total chirality, a dominant contribution comes from the layers without the monopoles. As one can see from layer $2_{xy}$ in Fig. \ref{fig:chidis_infieldHL} (b), $\chi^{\rm T, tetra}_z$'s on the layer without monopoles take nonzero values of the same sign, leading to a large total chirality. The existence of such a uniform chirality layer is characteristic of the hedgehog-lattice phase with topological objects of monopoles and antimonopoles.

In the chiral I phase having the orthorhombic-symmetric spin structure with the anisotropy originating from the double-${\bf Q}$ $S^z$ ordering in the plane perpendicular to the main symmetry axis, the total chirality $\mbox{\boldmath $\chi$}^{\rm T}$ is also nonzero (see the red regions in Figs. \ref{fig:Hdep} and \ref{fig:HTdep}). Figure \ref{fig:chidis_chiral1L} shows the layer-resolved chirality distribution in the chiral I phase, where each of all the layers has both positive and negative $\chi^{\rm T,tetra}_\alpha$'s. On the layers perpendicular to the main symmetry axis ($z$-axis), the positive and negative chiralities are completely cancelled out [see Fig. \ref{fig:chidis_chiral1L} (b)], whereas on the layers perpendicular to its orthogonal direction ($x$-axis), the cancellation is incomplete due to the population imbalance between the positive and negative chiralities, resulting in a nonzero total chirality $\chi^{\rm T}_x$ which takes a negative value in the case of Fig. \ref{fig:chidis_chiral1L} (a). The situation on the layers perpendicular to $y$-axis is the same as that for $x$-axis, so that the chirality vector $\mbox{\boldmath $\chi$}^{\rm T}$ is induced along the $[110]$ direction perpendicular to the main symmetry axis. The in-plane direction of $[110]$ is determined by the double-${\bf Q}$ ordering vectors of ${\bf Q}_1$ and ${\bf Q}_4$ satisfying the relation ${\bf Q}_1+{\bf Q}_4=(1,1,0)$. In a different case where ${\bf Q}_2$ and ${\bf Q}_4$ are picked up instead of ${\bf Q}_1$ and ${\bf Q}_4$, the total chiralities on the layers perpendicular to $x$- and $y$-axes take opposite signs, i.e., $\chi^{\rm T}_x=-\chi^{\rm T}_y$, so that $\mbox{\boldmath $\chi$}^{\rm T}$ is induced along the $[\overline{1}10]$ direction which corresponds to ${\bf Q}_2-{\bf Q}_3=(-1,1,0)$.  

Compared with the in-field hedgehog-lattice phase possessing the tetragonal-symmetric spin structure, the direction of the emergent chirality vector $\mbox{\boldmath $\chi$}^{\rm T}$ corresponding to the fictitious magnetic field is different. Furthermore, a uniform chirality layer such as layer $2_{xy}$ in Fig. \ref{fig:chidis_infieldHL} (b) does not appear in the chiral I phase.

In the case of the chiral I' phase shown in Fig. \ref{fig:chidis_chiral1H}, there exist positive and negative tetrahedron chiralities on most of the layers. In contrast to the chiral I phase, their populations are the same on each layer and their contributions are completely cancelled out, as is also confirmed from the reference calculation [see Eqs. (\ref{eq:tetra_chi_type1}) and (\ref{eq:tetra_chi_type3}) in Appendix B]. Thus, the total chirality does not emerge in the chiral I' phase. 

In the chiral II phase possessing the nonzero total chirality [see the green region in Fig. \ref{fig:Hdep} (b)], the situation is not so simple. Since the spin structure in the chiral II phase is orthorhombic, the chirality distributions on the layers perpendicular to the $x$-, $y$-, and $z$-axes are not equivalent, which can clearly be seen from Fig. \ref{fig:chidis_chiral2}. On the layer perpendicular to $z$-axis shown in Fig. \ref{fig:chidis_chiral2} (c), the local chirality is zero, as in the case of the high-field coplanar phase [see the lower panels in Fig. \ref{fig:chidis_coplanar} (c) in Appendix C]. Concerning the layers perpendicular to the $x$- and $y$-axes, although the populations of the positive and negative local chiralities are the same, a complete cancellation occurs only for the $y$-direction. On the layers perpendicular to $x$-axis shown in Fig. \ref{fig:chidis_chiral2} (a), the absolute values of positive and negative chiralities are slightly different, which results in a nonzero net contribution on each layer. The direction of $\mbox{\boldmath $\chi$}^{\rm T}$ is associated with the ordering vector at which the $(1,0,0)$-type Bragg peaks are absent [compare Figs. \ref{fig:chiral2} (b) and \ref{fig:chidis_chiral2}].

\section{Summary and discussion}
In this work, we have theoretically investigated the $J_1$-$J_3$ classical Heisenberg model on the breathing pyrochlore lattice without the DM interaction, putting emphasis on the stability of the hedgehog-lattice topological spin texture and the field-induced chirality $\mbox{\boldmath $\chi$}^{\rm T}$ associated with the chirality-driven anomalous Hall effect. In the model, the breathing lattice structure is characterized by the two different NN interactions $J_1$ and $J_1'$ which are defined on small and large tetrahedra, respectively. It is found by means of MC simulations that the hedgehog lattice characterized by the quadruple $(\frac{1}{2},\frac{1}{2},\frac{1}{2})$ magnetic Bragg reflections is stable irrespective of the signs of $J_1$ and $J_1'$ as long as $J_1 \neq J_1'$ and the third antiferromagnetic interaction along the bond direction $J_3$ is sufficiently strong. It is also found that in a magnetic field, there exist three chiral phases with nonzero field-induced chirality $\mbox{\boldmath $\chi$}^{\rm T} \neq 0$. They are the in-field hedgehog-lattice, chiral I, and chiral II phases in Fig. \ref{fig:HT_phaseall}, among which only the hedgehog lattice possesses the topological objects of the monopoles and antimonopoles. The direction of $\mbox{\boldmath $\chi$}^{\rm T}$ is determined by the symmetry of the spin structure. In particular, in the in-field hedgehog lattice with the tetragonal-symmetric spin structure, $\mbox{\boldmath $\chi$}^{\rm T}$ is induced along the tetragonal symmetry axis. 

We note that $\mbox{\boldmath $\chi$}^{\rm T}$ becomes nonzero only in the applied magnetic field, which is in sharp contrast to the associated two-dimensional analogue, a miniature skyrmion crystal (SkX) in a $J_1$-$J_3$ antiferromagnet on the breathing kagome lattice, where $\mbox{\boldmath $\chi$}^{\rm T}$ is nonzero even in the absence of the applied field and thus, a zero-field topological Hall effect is possible \cite{KagomeSkX_AK_22}.
Such a difference can be understood from the following fact: In the zero-field hedgehog lattice shown in Fig. \ref{fig:hedgehog_snapshot}, neighboring breathing-kagome-lattice layers stacking along the $(111)$ direction have opposite-sign uniform chiralities (compare spin configurations on equilateral triangles in neighboring kagome layers) and thus, the net chirality vanishes, whereas in the zero-field miniature SkX, such a cancellation does not occur due to the single-layer nature of the lattice. In a magnetic field, on the other hand, both the three-dimensional and two-dimensional spin textures commonly possess nonzero total chirality. 

It is useful to compare the present frustration-induced hedgehog lattice to the DM-induced hedgehog lattice in which the fictitious field corresponding to $\mbox{\boldmath $\chi$}^{\rm T}$  is also induced by the external magnetic field. In the DM case, the fictitious field emerges only along the applied field direction due to the field-induced position shifts of monopoles and antimonopoles \cite{Hedgehog_MFtheory_Park_11}, whereas in the present system, their positions are unchanged and $\mbox{\boldmath $\chi$}^{\rm T}$ emerges in any of the possible three directions $x$, $y$, and $z$, i.e., along the tetragonal symmetry axis of the spin structure. Furthermore, the sign of $\mbox{\boldmath $\chi$}^{\rm T}$ can be positive or negative reflecting the fact that the spin Hamiltonian (\ref{eq:Hamiltonian}) does not involve the DM interaction and thus, the right-handed and left-handed chiralities are degenerate as in other frustrated systems \cite{SkX_Okubo_12, SkX-RKKY_Mitsumoto_21, SkX-RKKY_Mitsumoto_22, KagomeSkX_AK_22}. Nevertheless, the DM-induced and present frustration-induced hedgehog lattices share a common feature in the origin of the field-induced total chirality. In the present system, the dominant contribution to $\mbox{\boldmath $\chi$}^{\rm T}$ comes from the uniform chirality layers sandwiched by the monopoles and antimonopoles, and in the DM system, it comes from skyrmion layers with a uniform chirality sandwiched by the position-shifted monopoles and antimonopoles \cite{MnGe_Kanazawa_16}. Such a uniform chirality layer does not appear in the chiral I and chiral II phases both of which do have nonzero $\mbox{\boldmath $\chi$}^{\rm T}$ but do not have topological structures. This suggests that the field-induced uniform chirality layer might be inherent to the topological objects of the monopoles and antimonopoles.

In experiments, the $(\frac{1}{2},\frac{1}{2},\frac{1}{2})$ spin correlation has not been observed in so-far reported breathing pyrochlore magnets such as the chromium oxides Li(Ga, In)Cr$_4$O$_8$ \cite{BrPyro_Okamoto_13,BrPyro_Tanaka_14,BrPyro_Nilsen_15,BrPyro_Saha_16,BrPyro_Lee_16,BrPyro_Saha_17,BrPyro_doped_Okamoto_15,BrPyro_doped_Wang_17,BrPyro_doped_Wawrzynczak_17,BrPyro_Hdep_Okamoto_17,BrPyro_Hdep_Gen_19} and sulfides Li(Ga, In)Cr$_4$S$_8$ \cite{BrPyro_Sulfides_Okamoto_18, BrPyro_Sulfides_Pokharel_18,BrPyro_Hdep_Gen_20,BrPyro_Sulfides_Kanematsu_20, BrPyro_Sulfides_Pokharel_20} and the quantum magnet Ba$_3$Yb$_2$Zn$_5$O$_{11}$ \cite{qBrPyro_Kimura_14, qBrPyro_Haku_prb16, qBrPyro_Haku_jpsj16, qBrPyro_Rau_16,qBrPyro_Rau_18}. In the uniform pyrochlore antiferromagnets Ge$B_2$O$_4$ ($B$=Ni, Co, Fe, Cu) \cite{GeNiO_Crawford_03,GeNiCoO_Diaz_06,GeNiO_Lancaster_06,GeNiO_Matsuda_08,GeCoO_Watanabe_08,GeCoO_Watanabe_11,GeCoO_Tomiyasu_11,GeAO_Barton_14,GeCoO_Fabreges_17,GeCoO_Pramanik_19,GeCuO_Zou_16}, on the other hand, the $(\frac{1}{2},\frac{1}{2},\frac{1}{2})$ magnetic LRO has been reported, although the experimentally proposed spin structure seems to be different from the ones in the present theoretical model. 
If one can modify exchange interactions in the above possible parent compounds, the $(\frac{1}{2},\frac{1}{2},\frac{1}{2})$ hedgehog lattice might be realized. Considering that the pyrochlore lattice has kagome-lattice layers as a building block, the kagome-lattice antiferromagnet BaCu$_3$V$_2$O$_8$(OD)$_2$ \cite{CoplanarOct_Boldrin_prl_18} might provide a useful information to realize a relatively strong $J_3$, since in this compound, $J_3$ is sufficiently strong and the coplanar ordered state could be well described by the two-dimensional version of the present model \cite{KagomeSkX_AK_22}.

Although in the above listed magnets are insulators, when the system can be tuned to be metallic, the chirality-driven anomalous Hall effect, which gradually increases with increasing field, should be observed. 
In a metallic system, the isotropic exchange spin interactions in Eq. (\ref{eq:Hamiltonian}) can be mediated by conduction electrons in the form of the Ruderman–Kittel–Kasuya–Yosida (RKKY) interaction, so that a conventional Kondo lattice model without spin-orbit interactions could be a minimum microscopic model. Then, the spin chirality is reflected in electron transport via the Kondo coupling \cite{THE_Tatara_02}. Since our result presented here is relatively robust against the second NN interaction $J_2$ (at least for AFM $J_1$ and $J_1'$ \cite{hedgehog_AK_prb_21}), the RKKY interaction which can effectively be mapped onto Eq. (\ref{eq:Hamiltonian}) could be realized by controlling the conduction electron density or the Fermi level.
Even in insulating systems, a thermal Hall effect might serve as a probe to detect chiral orders as in a DM system \cite{ThermalHall_Yamashita}, but we will leave this issue for our future work.   

\begin{acknowledgments}
The authors are thankful to ISSP, the University of Tokyo and YITP, Kyoto University for providing us with CPU time. This work is supported by JSPS KAKENHI Grant Number JP17H06137 and JP21K03469.
\end{acknowledgments}

\appendix
\section{Chirality structure factor}
\begin{figure}[t]
\begin{center}
\includegraphics[width=\columnwidth]{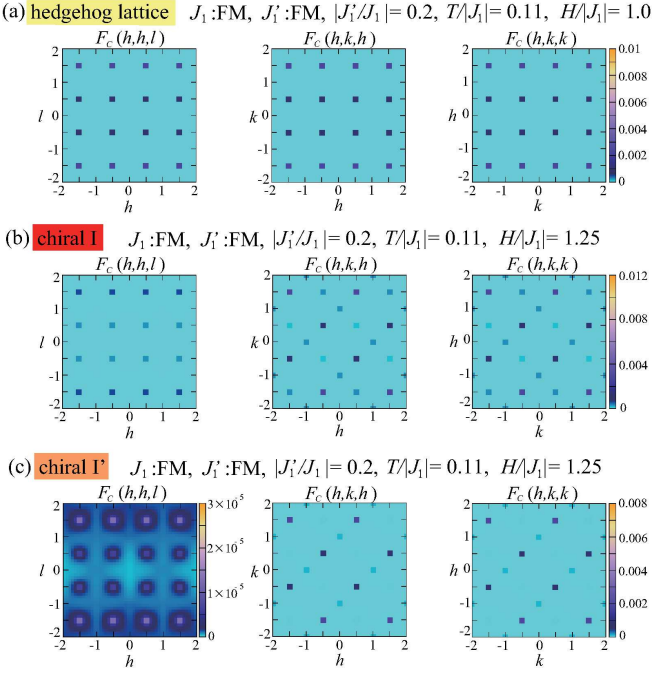}
\caption{Chirality structure factors $F_C({\bf q})$ in the $(h,h,l)$, $(h,k,h)$, and $(h,k,k)$ planes (from left to right) obtained in the MC simulations for $J_3/|J_1|=0.7$. (a), (b), and (c) are associated with Figs. \ref{fig:infieldHL}, \ref{fig:chiral1L}, and \ref{fig:chiral1H}, respectively. 
\label{fig:Cq2comp}}
\end{center}
\end{figure}
Figure \ref{fig:Cq2comp} shows the chirality structure factors $F_C({\bf q})$ in the in-field hedgehog-lattice, chiral I, and chiral I' phases. In the hedgehog-lattice phase, the quadruple-${\bf Q}$ $(\frac{1}{2},\frac{1}{2},\frac{1}{2})$ Bragg peaks have almost the same intensity, whereas in the chiral I phase, $(\frac{1}{2},\frac{1}{2},\frac{1}{2})$ and $(\frac{1}{2},\frac{1}{2},-\frac{1}{2})$ intensities are relatively weak. As readily seen from the left panels of Figs. \ref{fig:Cq2comp} (b) and (c), these intensities eventually vanish in the chiral I' phase, showing a double-${\bf Q}$ structure in the chirality sector. 
In the canted collinear, canted coplanar, and chiral II phases, the local chirality $\chi({\bf R}_l) = \sum_{i,j,k \in l \, {\rm th} \, {\rm tetra}} \chi_{ijk}$ vanishes, so that the associated $F_C({\bf q})$ [see Eq. (\ref{eq:F_C})] does not show any Bragg peaks (not shown in Fig. \ref{fig:Cq2comp}). In Fig. \ref{fig:Cq2comp}, one notices that a uniform ${\bf q}=0$ component is absent in $F_C({\bf q})$. Although at first sight, this may look inconsistent with the fact that in the hedgehog-lattice and chiral I phases, the total chirality $\mbox{\boldmath $\chi$}^{\rm T}$, i.e., a uniform chirality component, is nonzero, $F_{C}({\bf q})$ is not directly associated with the vector quantity $\mbox{\boldmath $\chi$}^{\rm T}$, as $F_{C}({\bf q})$ does not involve the geometrical factor $\hat{n}_{ijk}$ appearing in $\mbox{\boldmath $\chi$}^{\rm T}$ [see Eq. (\ref{eq:total_chi})].

\section{Reference calculation of chirality for a tetrahedron}
\begin{figure}[t]
\begin{center}
\includegraphics[scale=0.57]{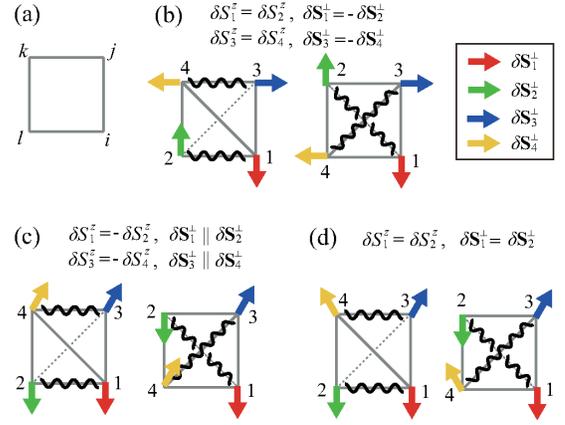}
\caption{Schematically drawn tetragonal-symmetric spin configurations on a tetrahedron projected onto a two-dimensional plane. (a) Definition of four sites on the projected tetrahedron, and (b)-(d) three typical spin configurations in the projected planes parallel (left) and perpendicular (right) to the tetragonal symmetry axis, where red, green, blue, and yellow arrows represent $S^xS^y$ components of spins $\delta {\bf S}^\perp_i$ ($i$=1-4) belonging to the sublattice 1, 2, 3, and 4, and other notations are the same as those in Fig. \ref{fig:tetraconf}. In (b) [(c)], the $S^xS^y$ components of paired spins $\delta {\bf S}^\perp_i$'s are antiparallel (parallel), whereas associated $S^z$ ones $\delta S^z_i$'s are parallel (antiparallel). In (d), two spins for one pair (sublattices 1 and 2) are pointing in the same direction. \label{fig:tetraconf_typical}}
\end{center}
\end{figure}
We examine the chirality vector $\mbox{\boldmath $\chi$}^{\rm T,tetra}$ for a tetragonal-symmetric spin configuration on a tetrahedron projected onto a two-dimensional plane.
Suppose that spins on the four corners of a tetrahedron be ${\bf S}_1$, ${\bf S}_2$, ${\bf S}_3$, and ${\bf S}_4$. As a uniform magnetization $m$ is induced by a magnetic field, these spins are expressed as
\begin{equation}
{\bf S}_i = (m + \delta S^z_i)\hat{z} + \delta {\bf S}^\perp_i.
\end{equation}
Then, the chirality for the projected tetrahedron shown in Fig. \ref{fig:tetraconf_typical} (a) is calculated as
\begin{eqnarray}\label{eq:tetra_chi_typical}
&&\chi_{ijk}+\chi_{jkl}+\chi_{kli}+\chi_{lij} \nonumber\\
&&=\delta S^z_i \, \hat{z}\cdot\big[  (\delta {\bf S}^\perp_j \times \delta {\bf S}^\perp_k) +  (\delta {\bf S}^\perp_k \times \delta {\bf S}^\perp_l) -  (\delta {\bf S}^\perp_l \times \delta {\bf S}^\perp_j)  \big] \nonumber\\
&&+\delta S^z_j \, \hat{z}\cdot\big[  (\delta {\bf S}^\perp_k \times \delta {\bf S}^\perp_l) +  (\delta {\bf S}^\perp_l \times \delta {\bf S}^\perp_i) -  (\delta {\bf S}^\perp_i \times \delta {\bf S}^\perp_k)  \big] \nonumber\\
&&+\delta S^z_k \, \hat{z}\cdot\big[  (\delta {\bf S}^\perp_l \times \delta {\bf S}^\perp_i) +  (\delta {\bf S}^\perp_i \times \delta {\bf S}^\perp_j) -  (\delta {\bf S}^\perp_j \times \delta {\bf S}^\perp_l)  \big] \nonumber\\
&&+\delta S^z_l \, \hat{z}\cdot\big[  (\delta {\bf S}^\perp_i \times \delta {\bf S}^\perp_j) +  (\delta {\bf S}^\perp_j \times \delta {\bf S}^\perp_k) -  (\delta {\bf S}^\perp_k \times \delta {\bf S}^\perp_i)  \big] \nonumber\\
&&+ 2m \, \hat{z}\cdot \big[  (\delta {\bf S}^\perp_i - \delta {\bf S}^\perp_k) \times  (\delta {\bf S}^\perp_j - \delta {\bf S}^\perp_l)  \big].
\end{eqnarray}
In the tetragonal-symmetric spin structures appearing in the present model, the four sublattices are paired up into two on each tetrahedron. As discussed in Sec. VI A, the paired sublattices are arranged in parallel (diagonally) on the tetrahedron projected onto a plane perpendicular (parallel) to the tetragonal symmetry axis (see Fig. \ref{fig:tetraconf}). We will discuss how such a difference in the arrangement pattern is reflected in $\mbox{\boldmath $\chi$}^{\rm T,tetra}$ for three typical tetrahedral spin configurations illustrated in Figs. \ref{fig:tetraconf_typical} (b)-(d). In Fig. \ref{fig:tetraconf_typical} (b), it is assumed for simplicity that the $S^xS^y$ components of paired-sublattice spins $\delta {\bf S}^\perp_i$ 's are antiparallel, whereas the $S^z$ components $\delta S^z_i$ 's are parallel. This type appears in the in-field hedgehog-lattice, chiral I', and low-field coplanar phases [see the tetrahedra enclosed by a green box in Figs. \ref{fig:infieldHL} (c), \ref{fig:chiral1H} (c), and \ref{fig:cantcoplanar} (c)]. Figure \ref{fig:tetraconf_typical} (c) illustrates the opposite case, i.e., $\delta {\bf S}^\perp_i$ 's for each sublattice pair are parallel, whereas $\delta S^z_i$ 's are antiparallel, which is realized in the in-field hedgehog-lattice phase  [see the tetrahedra enclosed by a purple box in Fig. \ref{fig:infieldHL} (c)]. The last type shown in Fig. \ref{fig:tetraconf_typical} (d) appears in the chiral I' and high-field coplanar phases [see Figs. \ref{fig:chiral1H} (c) and \ref{fig:cantcoplanar} (d)], where two spins in one pair are pointing in the same direction.
 
\begin{figure*}[t]
\begin{center}
\includegraphics[scale=0.57]{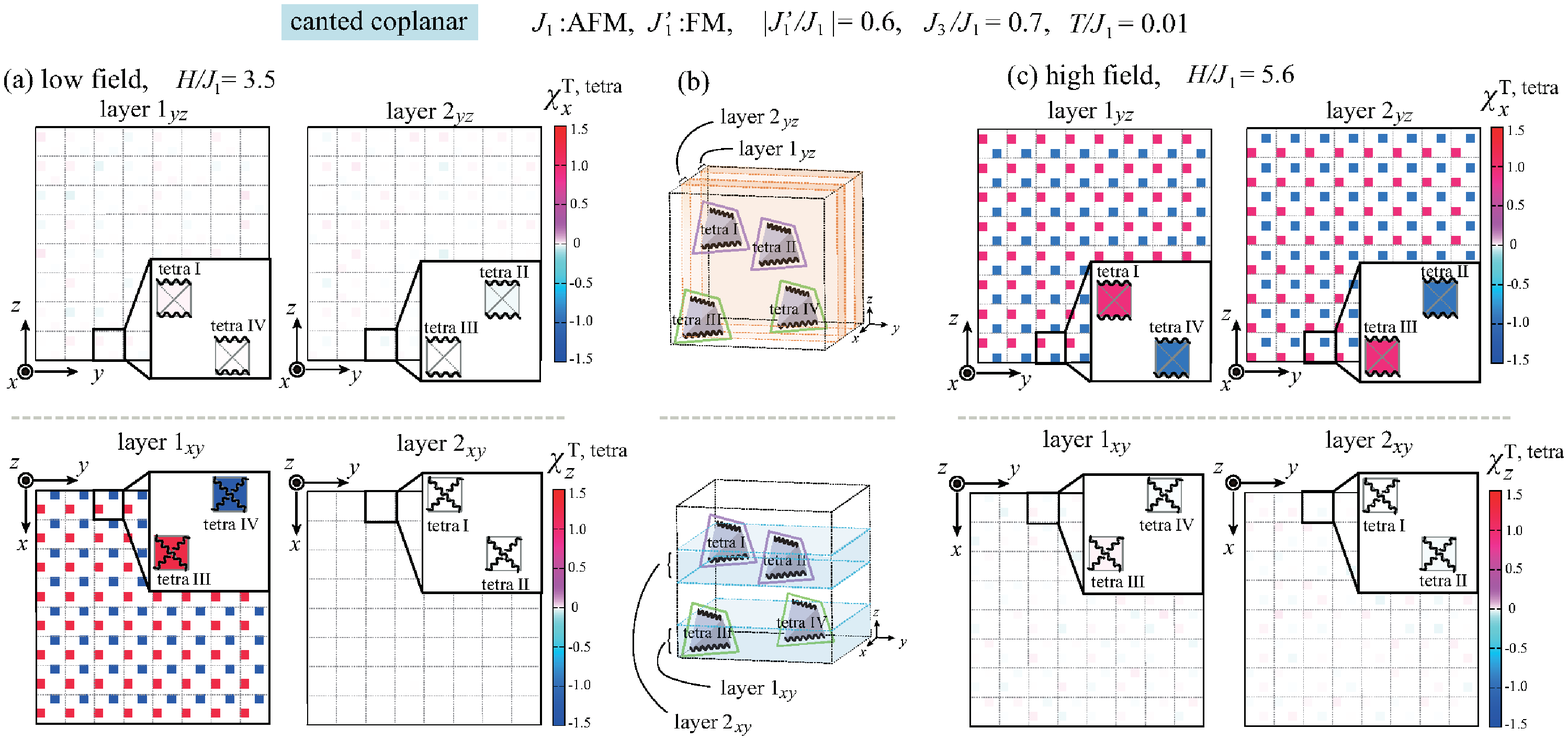}
\caption{Chirality distributions in the projected two-dimensional planes in the canted coplanar phase obtained at $T/J_1=0.01$ for $|J_1'/J_1|=0.6$ and $J_3/J_1=0.7$ with AFM $J_1$ and FM $J_1'$. (a) and (c) Layer-resolved distributions of the chirality calculated from the MC snapshots at the low field of $H/J_1=3.5$ and the high field of $H/J_1=5.6$ shown in Figs. \ref{fig:cantcoplanar} (c) and (d), respectively, where the layers are defined in (b). In (a) and (c), the upper (lower) panels show $\chi^{\rm T,tetra}_x$'s ($\chi^{\rm T,tetra}_z$'s) on the $yz$- ($xy$-)plane layers, where short-time average over 200 MC steps has been made to reduce the thermal noise. Notations of the paired sublattices and tetra's are the same as those in Figs. \ref{fig:cantcoplanar} (e) and (f). \label{fig:chidis_coplanar}}
\end{center}
\end{figure*}
For the spin configuration in Fig. \ref{fig:tetraconf_typical} (b), one can easily calculate the chirality vector by using Eq. (\ref{eq:tetra_chi_typical}) as follows:
\begin{equation}\label{eq:tetra_chi_type1}
\mbox{\boldmath $\chi$}^{\rm T,tetra}_{\rm (b)} = \left( \begin{array}{c}
0 \\
0 \\
4(2m+\delta S^z_1+\delta S^z_3) \, \hat{z}\cdot(\delta {\bf S}^\perp_1\times \delta {\bf S}^\perp_3)
\end{array} \right) ,
\end{equation}
where the relations $\delta S^z_1=\delta S^z_2$, $\delta {\bf S}^\perp_1 = -\delta {\bf S}^\perp_2$, $\delta S^z_3=\delta S^z_4$, and $\delta {\bf S}^\perp_3 = -\delta {\bf S}^\perp_4$ have been used and the trivial prefactor $1/\sqrt{3}$ has been omitted for simplicity. $\mbox{\boldmath $\chi$}^{\rm T,tetra}_{\rm (b)}$ is nonvanishing only in the tetragonal-symmetric $z$ direction. In other words, the chirality on the projected two dimensional plane is nonzero for the diagonal sublattice pairing, whereas zero for the parallel pairing. It should be noted that  the nonzero component $\chi^{\rm T,tetra}_{\rm (b), \, z}$ is an odd function with respect to $\delta {\bf S}^\perp_3$, so that it can be cancelled out by a counter tetrahedron in which $\delta {\bf S}^\perp_3$ and $\delta {\bf S}^\perp_4$ are interchanged. Such a cancellation actually occurs in the in-field hedgehog-lattice, chiral I', and low-field coplanar phases [see Figs. \ref{fig:chidis_infieldHL}, \ref{fig:chidis_chiral1H}, and \ref{fig:chidis_coplanar} (a)].

In the case of Fig. \ref{fig:tetraconf_typical} (c), the additional constraint $\delta S^z_1=\delta S^z_3$ for the in-field hedgehog lattice shown in Fig. \ref{fig:infieldHL} (c) yields
\begin{eqnarray}\label{eq:tetra_chi_type2}
&&\mbox{\boldmath $\chi$}^{\rm T,tetra}_{\rm (c)} = \left( \begin{array}{c}
-f^{(o)}(\delta S^z_1) \, \hat{z}\cdot(\hat{\delta S^\perp_1}\times \hat{\delta S^\perp_3}) \\
0 \\
f^{(e)}(\delta S^z_1) \, \hat{z}\cdot(\hat{\delta S^\perp_1}\times \hat{\delta S^\perp_3})
\end{array} \right) , \\
&& f^{(o)}(x) = 2x\big[4m^2+\big( \sqrt{1-(m+x)^2}+\sqrt{1-(m-x)^2}\big)^2 \big] , \nonumber\\
&& f^{(e)}(x) = 2m\big[4x^2+\big( \sqrt{1-(m+x)^2}-\sqrt{1-(m-x)^2}\big)^2 \big] , \nonumber
\end{eqnarray}
where the fixed spin-length constraint $|\delta {\bf S}^\perp_i|^2+(m+\delta S^z_i)^2=1$ has been used together with the relations $\delta S^z_1=-\delta S^z_2$, $\delta {\bf S}^\perp_1 = \delta {\bf S}^\perp_2$, $\delta S^z_3=-\delta S^z_4$, and $\delta {\bf S}^\perp_3 = \delta {\bf S}^\perp_4$. Although the $x$ component in Eq. (\ref{eq:tetra_chi_type2}) is nonvanishing, it can be canceled out by a counter contribution as $f^{(o)}(\delta S^z_1)$ is an odd function of $\delta S^z_1$. In contrast, $f^{(e)}(\delta S^z_1)$ involves an even function part, so that $\chi^{\rm T,tetra}_{\rm (c), \, z}$ can survive even after the summation over the projected two dimensional plane where the paired sublattices are arranged diagonally.  

The chirality vector for Fig. \ref{fig:tetraconf_typical} (d) is obtained in the same manner as those for Eqs. (\ref{eq:tetra_chi_type1}) and (\ref{eq:tetra_chi_type2}) as
\begin{eqnarray}\label{eq:tetra_chi_type3}
&&\mbox{\boldmath $\chi$}^{\rm T,tetra}_{\rm (d)} = \left( \begin{array}{c}
2 \, {\bf S}_1\cdot({\bf S}_3\times {\bf S}_4) \\
2 \, {\bf S}_1\cdot({\bf S}_3\times {\bf S}_4) \\
0
\end{array} \right) , \\
&& {\bf S}_1\cdot({\bf S}_3\times {\bf S}_4) = (m+\delta S^z_1)\hat{z}\cdot(\delta {\bf S}^\perp_3\times \delta {\bf S}^\perp_4)  \nonumber\\
&& +(m+\delta S^z_4)\hat{z}\cdot(\delta {\bf S}^\perp_1\times \delta {\bf S}^\perp_3)-(m+\delta S^z_3)\hat{z}\cdot(\delta {\bf S}^\perp_1\times \delta {\bf S}^\perp_4). \nonumber
\end{eqnarray}
The tetrahedron chirality in the projected two-dimensional plane vanishes along the tetragonal symmetry axis, whereas it takes the same nonzero value in the remaining two non-symmetric directions. Although the $x$ and $y$ components of $\mbox{\boldmath $\chi$}^{\rm T,tetra}_{\rm (d)}$ are nonvanishing at the level of the single tetrahedron, these contributions proportional to ${\bf S}_1\cdot({\bf S}_3\times {\bf S}_4)$ can be completely canceled out in the following reason. When $\delta S^z_3=\delta S^z_4$ like in the case of the high-field coplanar phase, ${\bf S}_1\cdot({\bf S}_3\times {\bf S}_4)$ is an odd function with respect to the interchange between $\delta {\bf S}^\perp_3$ and $\delta {\bf S}^\perp_4$, so that it may be canceled out by a neighboring counter contribution. As will be demonstrated in Appendix C, this is actually the case for the high-field coplanar phase.

\section{Real-space chirality distributions in the canted coplanar phase}
Figure \ref{fig:chidis_coplanar} shows the layer-resolved real-space chirality distributions in the canted coplanar phase. For the low-field coplanar structure shown in Fig. \ref{fig:cantcoplanar} (c) , the chirality vectors $\mbox{\boldmath $\chi$}^{\rm T, tetra}$'s on tetra's I and II are trivially absent and the ones on tetra's III and IV are nonvanishing only for the tetragonal-symmetric axis, which can clearly be seen in Fig. \ref{fig:chidis_coplanar} (a). The nonvanishing components $\chi^{\rm T,tetra}_z$'s take positive and negative signs on each layer, being completely cancelled by each other [see Eq. (\ref{eq:tetra_chi_type1})], so that the total chirality vanishes. 

For the high-field coplanar structure shown in Fig. \ref{fig:cantcoplanar} (d), Eq. (\ref{eq:tetra_chi_type3}) suggests that the tetrahedron chirality $\mbox{\boldmath $\chi$}^{\rm T,tetra}$ are nonvanishing only for the non-symmetric directions ($x$- and $y$-directions), which can be seen in Fig. \ref{fig:chidis_coplanar} (b). Since the nonvanishing contributions on neghboring tetrahedra take opposite signs, they are completely cancelled out, resulting in no net chirality.

\end{document}